\documentclass[]{aiaa-tc}

\usepackage{graphicx}
\usepackage{epstopdf}
\usepackage{subfig}
\usepackage{amsmath, amssymb}
\usepackage{color, colortbl}
\usepackage{xcolor}
\usepackage{framed}
\usepackage{ulem}

 \title{Least Squares Shadowing for Sensitivity Analysis of Turbulent Fluid Flows}

 \author{
  Patrick J. Blonigan%
    \thanks{Graduate Student, Aeronautics and Astronautics, Cambridge, MA, AIAA Member.}
   , Steven A. Gomez\thanksibid{1}
  \ and Qiqi Wang 
  \thanks{Assistant Professor, Aeronautics and Astronautics, Cambridge, MA, AIAA Member.}\\
  {\normalsize\itshape
   Massachusetts Institute of Technology, Cambridge, MA, 02139, USA}
 }

 \AIAApapernumber{YEAR-NUMBER}
 \AIAAconference{SciTech, 2014, National Harbor, MD}
 \AIAAcopyright{\AIAAcopyrightD{2013}}

\newcommand{\dd}[2]{ \frac{d {#1}}{d {#2} } }

\newcommand{\pd}[2]{ \frac{\partial {#1}}{\partial {#2}} }

\newcommand{\pdt}[1]{ \pd{#1}{t} }
\newcommand{\pdu}[1]{ \pd{#1}{u} }
\newcommand{\mean}[1]{\bar{#1}}
\newcommand{\hf}{\frac{1}{2}}
\newcommand{\Uh}{\hat{U}(\mathbf{k},t)}
\newcommand{\Ph}{\hat{P}(\mathbf{k},t)}
\newcommand{\lpar}[1]{\left({#1}\right)}

\begin{document}

\maketitle

\begin{abstract}
Computational methods for sensitivity analysis are invaluable tools for aerodynamics research and engineering design.  However, traditional sensitivity analysis methods break down when applied to long-time averaged quantities in turbulent fluid flow fields, specifically those obtained using high-fidelity turbulence simulations. This is because of a number of dynamical properties of turbulent and chaotic fluid flows, most importantly high sensitivity of the initial value problem, popularly known as the ``butterfly effect''.
 
 The recently developed least squares shadowing (LSS) method avoids the issues encountered by traditional  sensitivity analysis methods by approximating the ``shadow trajectory'' in phase space, avoiding the high sensitivity of the initial value problem.  The following paper discusses how the least squares problem associated with LSS is solved.  Two methods are presented and are demonstrated on a simulation of homogeneous isotropic turbulence and the Kuramoto-Sivashinsky (KS) equation, a 4th order chaotic partial differential equation.  We find that while LSS computes fairly accurate gradients, faster, more efficient linear solvers are needed to apply both LSS methods presented in this paper to larger simulations.  
\end{abstract}

%

\section{Introduction}

Computational methods for sensitivity analysis are invaluable tools for fluid mechanics research and engineering design.  These methods compute derivatives of outputs with respect to inputs in computer simulations. However, traditional sensitivity analysis methods break down when applied to long-time averaged quantities in chaotic and turbulent fluid flow-fields, specifically those obtained using high-fidelity turbulence simulations. As many key scientific and engineering quantities of interest in turbulent flows are long-time averaged quantities, methods of computing their sensitivities are useful in design optimization \cite{Jameson1988}, data assimilation \cite{TELA:TELA459}, flow control \cite{Gunzburger:2002:PFC:640624} and uncertainty quantification \cite{uqreview}.  However, a number of dynamical properties of chaotic fluid flows, most importantly Edward Lorenz's ``Butterfly Effect'' make the formulation of robust and efficient sensitivity analysis methods difficult.  

The ``Butterfly Effect'' is a high sensitivity to initial conditions \cite{Lorenz:1963:det}. In chaotic flow fields, any small perturbation to the flow field will eventually result in drastic changes in the instantaneous flow-field. Consider a system of governing equations:  

\begin{equation}
\frac{\partial u}{\partial t} = f(u; s)
\label{e:dynSys}
\end{equation}

For a CFD simulation, this system would be the Navier-Stokes equations and the state variable $u$ would be a vector containing the velocity, pressure and temperature fields.  $s$ is some design parameter, such as the angle of attack of an airfoil.  Now consider some quantity of interest, $J$:

\[
J(t;s) = J(u(t;s))
\]

$J$ could be the instantaneous drag on an airfoil.  In many cases we are interested in the time-averaged quantities, so we define:

\[
\bar{J}(s) = \frac{1}{T} \int_0^T J(t;s)dt, \quad \bar{J}^{\infty}(s) = \lim_{T\to\infty} \bar{J}(s)
\]

\noindent Where $\bar{J}^{\infty}$ is the infinite time average.  For many non-chaotic systems, $\bar{J}$ will converge to $\bar{J}^{\infty}$.  However, this is not the case for chaotic systems, for which:

\begin{equation}
\frac{d\bar{J}^{\infty}}{ds} \neq  \lim_{T\to\infty} \frac{d\bar{J}}{ds}
\label{e:noComm}
\end{equation}

In fact, the difference between these two derivatives often grows as $T$ is increased \cite{Lea:2000:climate_sens}.  This is because the tangent (or adjoint) solution diverges exponentially, due to one of the system's unstable mode associated with its positive Lyapunov exponent.  As a result, the derivatives with respect to $s$ in the limit of $T\rightarrow\infty$ in equation (\ref{e:noComm}) do not commute.  

Prior work in sensitivity analysis of chaotic dynamical systems and fluid flows has been done almost exclusively by the meteorological community.  This work include the ensemble-adjoint method proposed by Lea et al. \cite{Lea:2000:climate_sens}. This method has been applied to an ocean circulation model with some success \cite{Lea:2002:ocean}.  Eyink et al. then went on to generalize the ensemble-adjoint method \cite{Eyink:2004:ensmbl}.  The ensemble-adjoint method involves averaging over a large number of ensemble calculations.  It was found by Eyink et al. that the sample mean of sensitivities computed with the ensemble adjoint approach only converges as $N^{-0.46}$, where $N$ is the number of samples, making it less computationally efficient than a naive Monte-Carlo approach \cite{Eyink:2004:ensmbl}.  

More recently, a Fokker-Planck adjoint approach for climate sensitivity 
analysis has been derived \cite{Thuburn:2005:FP}.  This approach involves 
finding a probability density function which satisfies a Fokker-Planck equation 
to model the climate.  The adjoint of this Fokker-Planck equation is then used 
to compute derivatives with respect to long time averaged quantities.  However, this method requires the discretization of phase space, making it computationally infeasible for systems with a high number of dimensions.  Also, the method requires adding diffusion into the system, potentially making the computed sensitivities inaccurate.  Methods based on fluctuation dissipation theorem (FTD) has been successful in computing climate sensitivities for complex chaotic dynamical systems \cite{0951-7715-20-12-004}.  However, for strongly dissipative systems whose SRB measure \cite{youngSRB} deviates strongly from Gaussian, fluctuation dissipation theorem based methods can be inaccurate.

A new method developed by the authors of this paper, the least squares shadowing (LSS) method, avoids these issues and is able to compute sensitivities of interest for chaotic systems and therefore turbulent flows by using the shadowing lemma \cite{Pilyugin:1999:shadow} and assuming ergodicity; that the long time averaged behavior of the system is independent of the initial condition.  

LSS can enable gradient-based optimization of systems with complicated, turbulent flows.  With the ever increasing power of modern computers, potential applications of LSS include the design of lifting bodies, fuel injection systems in jet engines or scramjet combustors, and rocket engines.  Overall, if LSS can be efficiently scaled to large scale CFD problems, it can lead to new computational design method which could have a potentially large impact on the aerospace industry.   

The following paper presents the first successful application of LSS to a large scale CFD problem, as well as a new LSS algorithm tailored to solve large scale problems.  First the two key concepts underpinning LSS, Lyapunov exponents and the shadowing lemma, are discussed.  Then two algorithms for LSS are presented and demonstrated. The first algorithm is demonstrated on a homogeneous isotropic turbulence simulation, the second algorithm is demonstrated on the Kuramoto-Sivashinsky equation, a 4th order, chaotic PDE.  Additionally, the advantages and disadvantages of both methods are discussed.  We conclude this paper with a discussion of current and future research directions for LSS.  

\section{Least Squares Shadowing}
\label{c:LSS}

Before we begin our discussion of LSS and its underpinning mathematical principles, it is helpful to review Lyapunov exponents and Lyapunov covariant vectors.  For some system $\frac{du}{dt} = f(u)$, there exist Lyapunov covariant vectors $\phi_1(u),\phi_2(u),...,\phi_i(u)$ corresponding to each Lyapunov exponent $\Lambda_i$, which satisfy the equation \cite{Ginelli:2007:Lya}:
\[
\frac{d}{dt}\phi_i(u(t)) = \left. \frac{\partial f}{\partial u} \right|_{u(t)} \cdot \phi_i(u(t)) - \Lambda_i \phi_i(u(t))
\]

\begin{figure}
\centering
\includegraphics[width=0.75\textwidth]{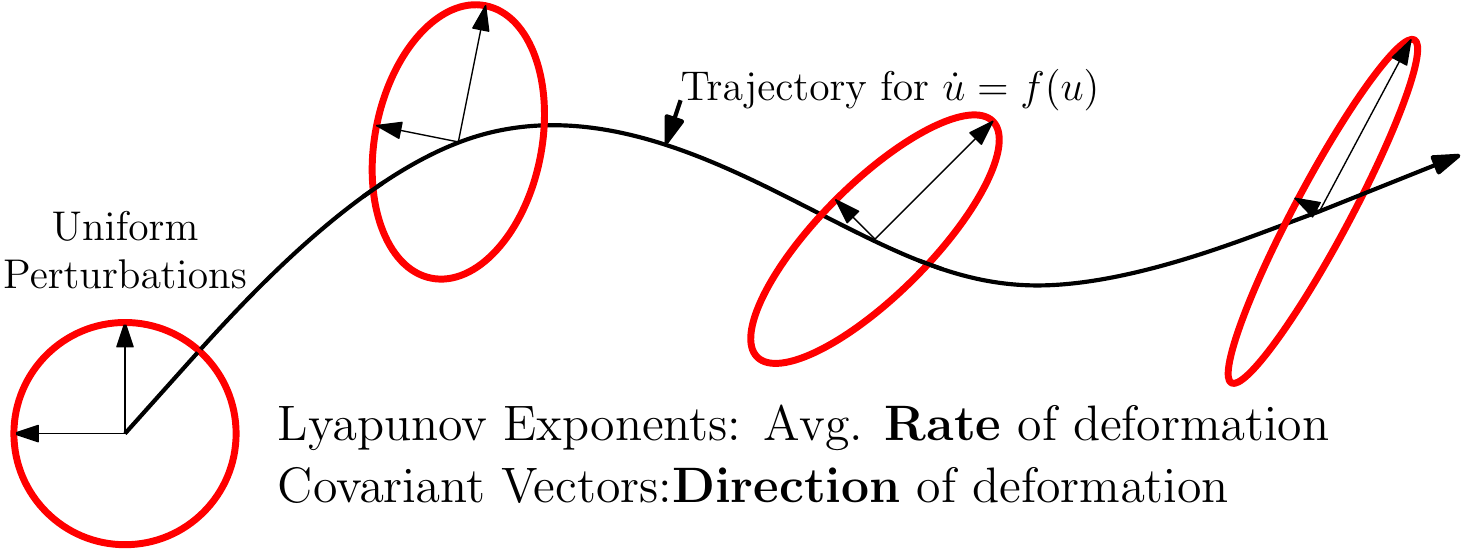}
\caption{Schematic of Lyapunov exponents and covariant vectors}
\label{f:Lyapunov}
\end{figure}

To understand what $\Lambda_i$ and $\phi_i$ represent, consider a sphere comprised of perturbations $\delta f$ to a system $\frac{\partial u}{\partial t} = f(u)$ at some time, as shown in the far left of figure \ref{f:Lyapunov}. As this system evolves in time, this sphere expands in some directions, contracts in some, and remains unchanged in others.  The rate at which the sphere expands or contracts corresponds to the Lyapunov exponent $\Lambda_i$ and the corresponding direction of expansion or contraction is the Lyapunov covariant vector $\phi_i$.  It is important to note that the $\phi_i$ are not the same as local Jacobian eigenvectors, each $\phi_i$ depends on all Jacobians along a trajectory. Also, the $\phi_i$ are not necessarily orthogonal, but the number of Lyapunov covariant vectors is the same as the number of dimensions of the system.  

\begin{figure}
\centering
\includegraphics[width=0.9\textwidth]{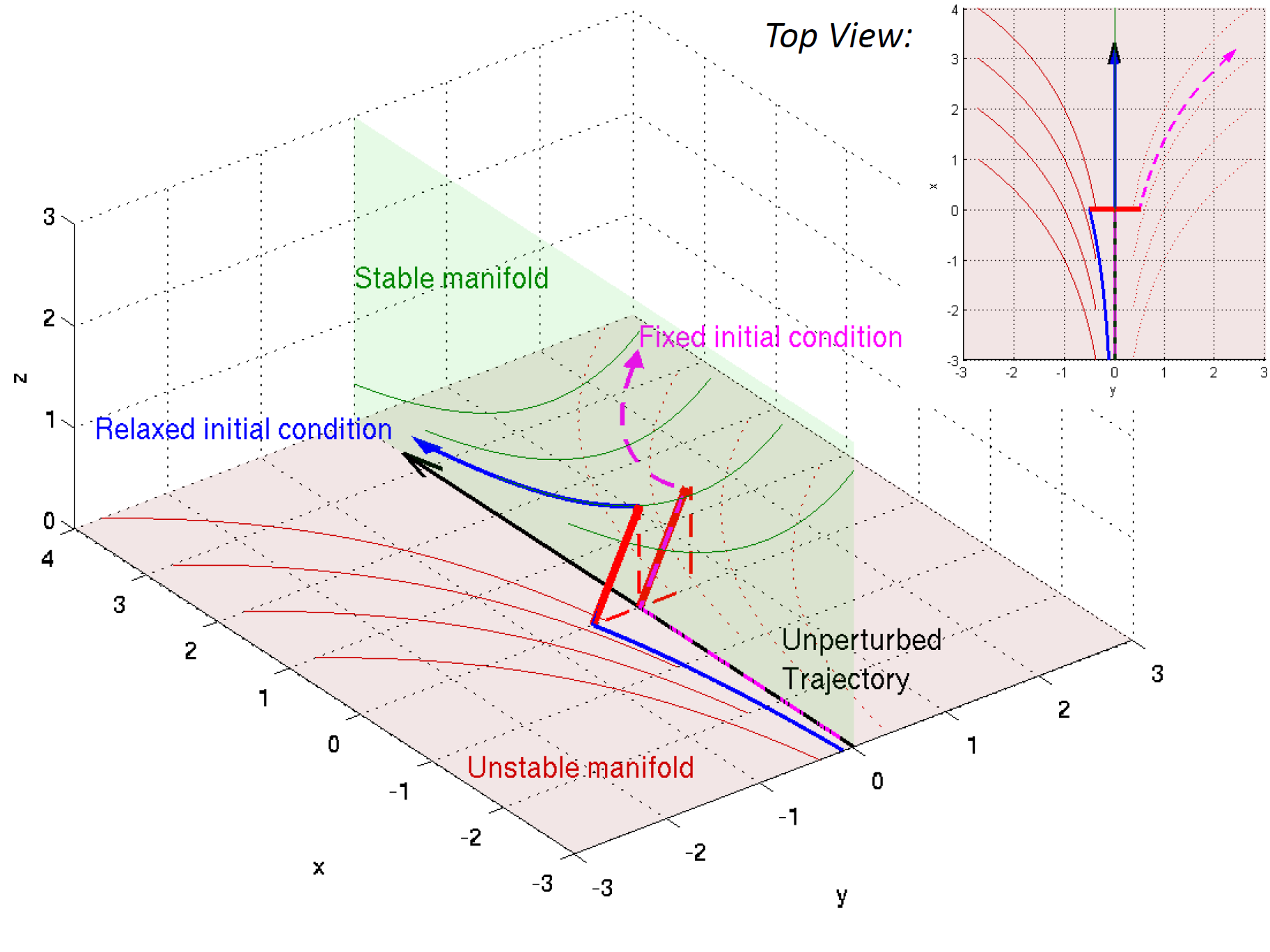}
\caption{Phase space trajectory of a chaotic dynamical system.  The unstable manifold, in red, is the space of all Lyapunov covariant vectors corresponding to positive exponents. The stable manifold, in green, corresponds to the space of all covariant vectors associated with negative exponents.  A perturbation to the system (in red) has components in both manifolds, and the unstable component causes the perturbed trajectory (pink) to diverge exponentially from the unperturbed trajectory (in black).  LSS chooses a perturbed trajectory with a different initial condition (in blue) that does not diverge from the unperturbed trajectory.  }
\label{f:LSS_schematic}
\end{figure}

A strange attractor, the type of attractor associated with dissipative chaotic dynamical systems, has at least one positive and one zero Lyapunov exponent \cite{Ginelli:2007:Lya}.  To illustrate the effect of the positive exponent, we consider figure \ref{f:LSS_schematic}.  We see that if the perturbed trajectory has the same initial condition as the unperturbed trajectory, the two trajectories diverge exponentially, leading to the issues with traditional sensitivity analysis discussed in the introduction.

However, the assumption of ergodicity means that it is not necessary to compare a perturbed and an unperturbed trajectory with the same initial condition if the quantities of interest are long time averages.  Therefore, an initial condition can be chosen such that the perturbed and unperturbed trajectories do not diverge, resulting in the blue trajectory in figure \ref{f:LSS_schematic}. The existence of this trajectory, called a ``shadow trajectory'', follows from the shadowing lemma \cite{Pilyugin:1999:shadow} :
{\it
Consider a reference solution $u_{ref}$ to equation \eqref{e:dynSys}.  If this system has a Hyperbolic strange attractor and if some system parameter $s$ is slightly perturbed: 

For any $\delta > 0$ there exists $\varepsilon > 0$, such that for every $u_{ref}$ that satisfies $\|d u_{ref} / d t - f(u_{ref})\|<\varepsilon$, $0 \le t \le T$, there exists a true solution $u_s$ and a time transformation $\tau(t)$, such that $\|u_s(\tau(t))-u_{ref}(t)\| < \delta$, $|1-d\tau/dt|<\delta$ and $d u_s/d \tau - f(u_s) = 0$, $0 \le \tau \le \mathcal{T}$. }\footnote{Note that $\|\cdot\|$ refers to distance in phase space}

Therefore, relaxing the initial condition allows us to find the shadow trajectory $u_s(\tau)$.  The key assumption of the shadowing lemma is that the attractor associated with the system of interest is {\em hyperbolic}.  The key property of hyperbolic attractors for the shadowing lemma is that tangent space can be decomposed into stable, neutrally stable, and unstable components everywhere on the attractor \cite{Hasselblatt:2002:hyperbolic}.  These components of tangent space correspond to negative, zero and positive Lyapunov exponents respectively and each component is spanned by the corresponding Lyapunov covariant vectors. Although the tangent space of many attractors cannot be decomposed into stable, neutral, and unstable components on the entire attractor, the {\em Chaotic Hypothesis} conjectures that many high-dimensional chaotic systems behave as if they were hyperbolic \cite{Eyink:2004:ensmbl}.  For example, the single non-hyperbolic point on the Lorenz attractor is the unstable fixed point at the origin.  Since most phase space trajectories do not pass through the unstable fixed point, the shadowing lemma holds for most trajectories on the Lorenz attractor.  

\begin{figure}
\centering
\includegraphics[width=0.35\textwidth]{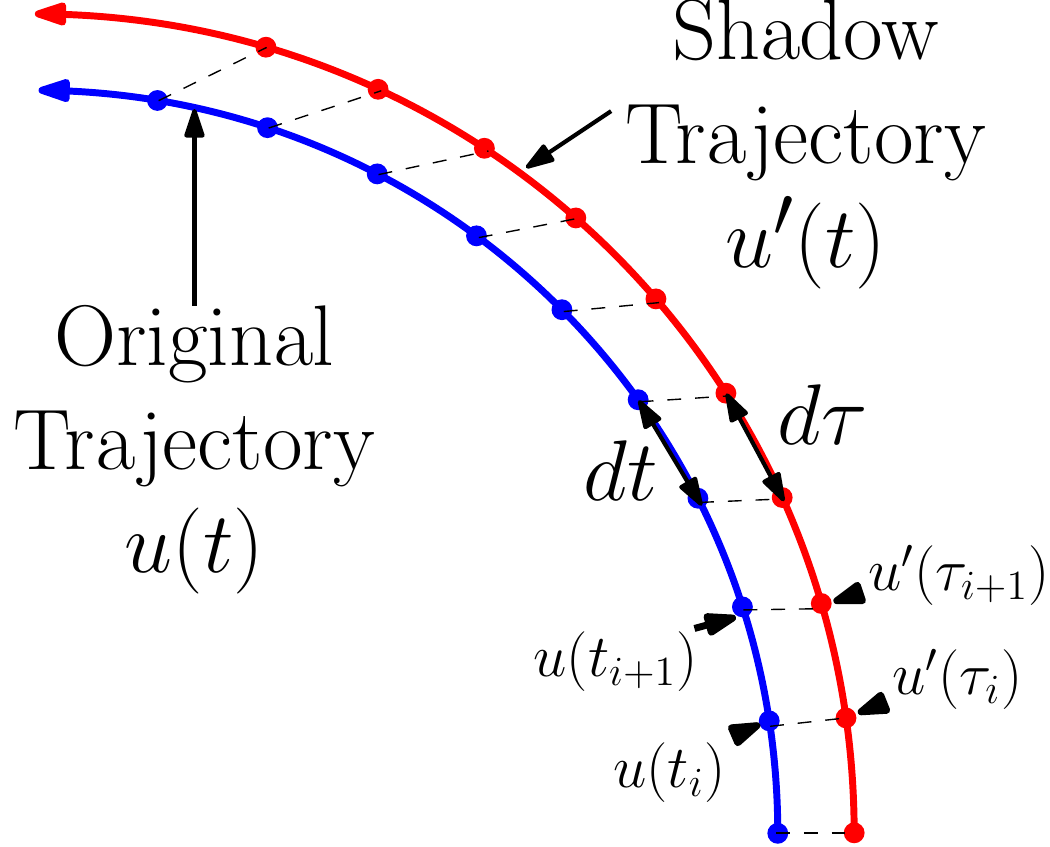}
\includegraphics[width=0.35\textwidth]{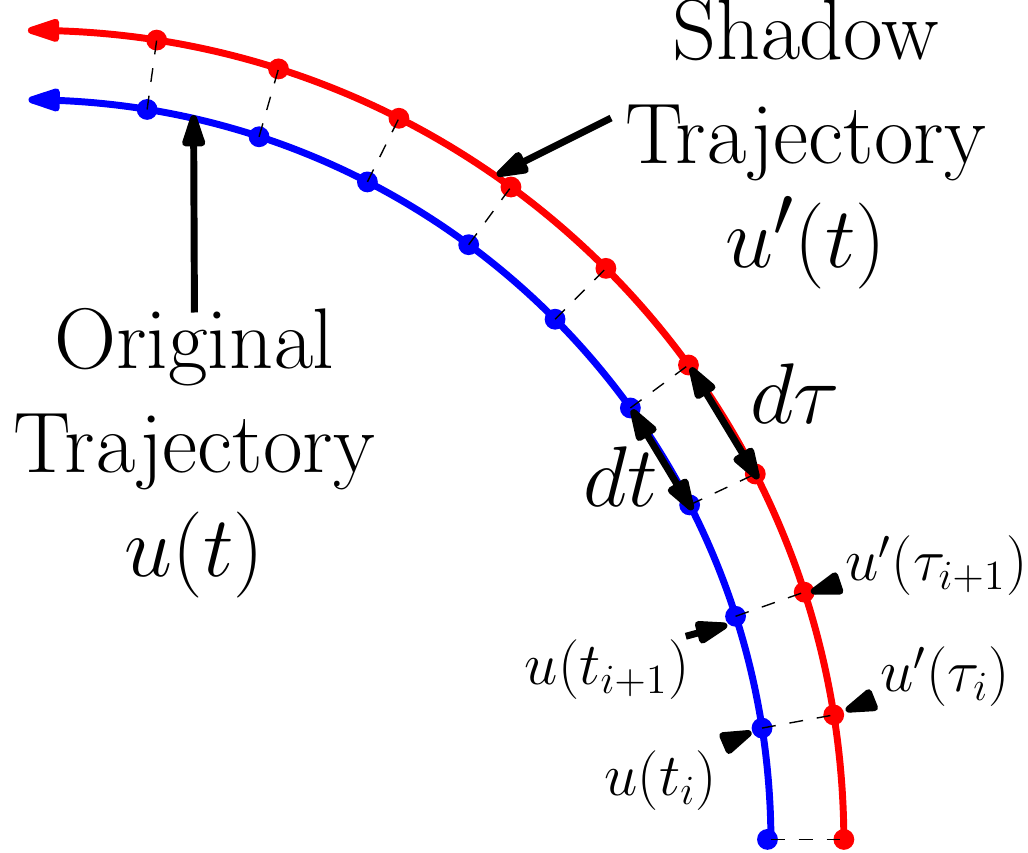}
\caption{LEFT: Original and shadow phase space trajectories without any time transformation ($d\tau/dt = 1$).  RIGHT: Original and shadow phase space trajectories with a time transformation $d\tau/dt = 1 + \eta$ that minimizes the distance between the two trajectories in phase space for all time.  }
\label{f:time_dilation}
\end{figure}

The time transformation alluded to in the shadowing lemma is required to deal with the zero (neutrally stable) Lyapunov exponent, whose covariant vector is simply $f(u)$.  The time transformation, referred to as ``time dilation'' in this paper and other LSS literature, is required to keep a phase space trajectory and its shadow trajectory close (in phase space) for all time, as shown in figure \ref{f:time_dilation}.

\section{Least Squares formulation}
\label{s:LSform}  

Although we could find a shadow trajectory by computing Lyapunov exponents and covariant vectors \cite{Wang:2013:LSS1}, this is not necessary.  Instead, a least squares problem is solved.  In this problem, we seek to minimize the L2 norm of the difference between the shadow trajectory $u$ and some reference solution $u_{ref}$ that satisfies $\dd{u_{ref}}{t} = f(u_{ref};s)$:

\begin{equation}
 \min_{u} \frac{1}{2} \int_{T_0}^{T_1} \| u(\tau(t)) - u_{ref}(t) \|^2 \ dt, \quad \text{s.t.} \ \dd{u}{\tau} = f(u;s+\delta s), \, T_0 < t < T_1
 \label{e:LeastSqs}
\end{equation}

The solution of equation \eqref{e:LeastSqs} is a solution $u(\tau)$ that remains close to $u_{ref}(t)$ in phase space from $T_0 < t < T_1$.  It can be shown that this $u(\tau)$ converges to the shadow trajectory as $T_1 \to \infty$ and $T_0 \to -\infty$\cite{Wang:2013:LSSthm}.  
For forward sensitivity analysis, we are interested in $v \equiv \partial u/\partial s$.  To find this, divide the minimization statement equation \eqref{e:LeastSqs} by $\delta s$ and take the limit $\delta s \to 0$ to obtain:

\begin{equation}
 \min_{v} \frac{1}{2} \int_{T_0}^{T_1} \| v(t) \|^2 \ dt, \quad \text{s.t.} \ \frac{dv}{dt} = \frac{\partial f}{\partial u} v + \frac{\partial f}{\partial s} + \eta f, \quad T_0 < t < T_1
 \label{e:LSSbasic}
\end{equation}

where $\eta \equiv d\tau/dt-1$ is the time dilation term, corresponding to the time transformation from the shadowing lemma discussed in the previous section.  

Equation \eqref{e:LSSbasic} is a linearly constrained least-squares problem, with the following KKT equations, derived using calculus of variations:

\begin{align}
\frac{\partial w}{\partial t} &= -\left(\frac{\partial f}{\partial u}\right)^* w - v, \quad
w(0) = w(T) = 0 \label{e:KKTw}\\
&\langle f, w \rangle = 0 \label{e:KKTeta}\\
\frac{dv}{dt} &= \frac{\partial f}{\partial u} v + \frac{\partial f}{\partial s} + \eta f \label{e:KKTv}
\end{align}

\noindent The LSS solution can be used to compute gradients for some quantity of interest $J$ (i.e. Drag):  
\begin{equation}
\frac{\partial \bar{J}}{\partial s} = \overline{\left\langle \frac{\partial J}{\partial u}, v \right\rangle} + \overline{\eta J} - \overline{\eta} \overline{J}
\label{e:LSSgrad}
\end{equation}
\noindent Where $\overline{x} \equiv \frac{1}{T} \int_0^T x \ dt$. See Wang et al., 2013 \cite{Wang:2013:LSS2} for a derivation of the equations \eqref{e:LSSbasic} to \eqref{e:LSSgrad} \footnote{Set $\alpha^2 = 0$ to obtain the equations presented in this paper}.  

There are several potential methods for solving equation \eqref{e:LSSbasic}.  
\begin{itemize}
 \item {\bf LSS Type I, Initial Condition Design:} We could search for the initial condition for the tangent equation, $v_0$, that solves \eqref{e:LSSbasic}.  This approach results in an optimization problem where the number of variables is equal to the number of states of the system (the size of $v_0$).  Starting from $v_0$ ensures that we satisfy the tangent condition, so LSS type I has no constraints.  However, the condition number of LSS type I scales with $e^{\Lambda T}$, where $\Lambda$ is the largest positive Lyapunov exponent of the system and $T=T_1-T_0$\cite{Blonigan:2013:LSS}.  Therefore, LSS type I becomes poorly conditioned for large values of $T$.  
\item {\bf LSS type II, Full Trajectory Design:} We could search the space of trajectories $v(t)$.  This approach results in a much better conditioned optimization problem (condition number scales with $T^2$)\cite{Blonigan:2013:LSS}.  However, each tangent solution $v(t)$ is very large.  For a $m$ time step simulation of a system with $n$ states, $v(t)$ has $mn$ elements.  Since $v(t)$ needs to satisfy the tangent equation \eqref{e:KKTv}, trajectory design has a constraint for each time step.  This means that type II involves the solution of a very large system of Karush-Kuhn-Tucker (KKT) equations.  
\item {\bf LSS type III, Checkpoint Design:} Instead of choosing an initial condition, or $v(t)$ for $T_0<t<T_1$ that solves \eqref{e:LSSbasic}, we could choose $v(t)$ at a few times $t_0,t_1,...,t_K$ between $T_0$ and $T_1$.   LSS type III is a compromise between type I and II.  It is better conditioned than type I and has fewer variables and constraints than type II.  Additionally, we will show that LSS type III has a relatively straight forward algorithm.  
\end{itemize}

The two latter approaches are discussed in more detail in the following sections.  

\section{LSS type II: Full Trajectory Design}
\label{s:typeII}

For implementations of LSS type II, we modify the least squares formulation for better numerical properties:

\begin{equation}
 \min_{v,\eta} \frac{1}{2}\int_{T_0}^{T_1} \| v \|^2 + \alpha^2 \| \eta \|^2 dt, \quad s.t. \quad \frac{dv}{dt} = \frac{\partial f}{\partial u} v + \frac{\partial f}{\partial s} + \eta f \,, \quad T_0<t<T_1 ,
\label{e:LSStypeII}
\end{equation}

Including $\eta$ in the minimization statement is consistent with the shadowing lemma, which states that $\|u(\tau(t))-u_{ref}(t)\| < \delta$ and $|1-d\tau/dt|=|\eta|<\delta$ for some $\delta$. $\alpha^2$ is simply a weighting factor, which has a considerable effect on the convergence of the numerical solver used for \eqref{e:LSStypeII} \cite{Blonigan:2013:MG, gomez:2013:masters}.  

\noindent The corresponding KKT system is \cite{Wang:2013:LSS2}:

\begin{align}
\frac{\partial w}{\partial t} &= -\left(\frac{\partial f}{\partial u}\right)^* w - v, \quad
w(0) = w(T) = 0 \label{e:KKTwII}\\
\alpha^2 \eta &= - \langle f, w \rangle \label{e:KKTetaII}\\
\frac{dv}{dt} &= \frac{\partial f}{\partial u} v + \frac{\partial f}{\partial s} + \eta f \label{e:KKTvII}
\end{align}

Equations (\ref{e:KKTwII}), (\ref{e:KKTetaII}) and (\ref{e:KKTvII}) can be combined to form a single second order equation for the Lagrange multiplier $w$\cite{Blonigan:2013:LSS}:

\begin{equation}
-\frac{d^2 w}{d t^2} - \left(\frac{d}{d t} \left(\frac{\partial f}{\partial u}\right)^*  - \frac{\partial f}{\partial u} \frac{d}{d t} \right) w + \left(\frac{\partial f}{\partial u}\left(\frac{\partial f}{\partial u}\right)^* +\frac{1}{\alpha^2} ff^*\right) w = \frac{\partial f}{\partial s}, \quad w(0) = w(T) = 0
\label{e:SPD_LSS}
\end{equation}

Equation (\ref{e:SPD_LSS}) shows that the LSS method has changed the tangent
equation from an initial value problem to a boundary value problem in time.   

To solve the system numerically, equations (\ref{e:KKTwII}), (\ref{e:KKTetaII}) and (\ref{e:KKTvII}) are discretized using finite differences and combined to form the following symmetric system:  

{\scriptsize
\[
\left( \begin{array}{ccccc|cccc|cccc}
\rowcolor{red!70} I & & & & & & & & & F_0^T & & & \\
\rowcolor{red!70} & I & & & & & & & & G_1^T & F_1^T & & \\
\rowcolor{red!70} & & \ddots & & & & &  & & & G_2^T & \ddots & \\
\rowcolor{red!70} & & & I & & & & & &   & & \ddots & F_{m-1}^T \\
\rowcolor{red!70} & & & & I & & & & & & & & G_m^T \\\hline
 & & & & & \alpha^2 & & & & f_1^T &  & &  \\
 & & & & & & \alpha^2 & & &  & f_2^T &  &  \\
 & & & & & & & \ddots & & & & \ddots & \\
 & & & & & & & & \alpha^2 &  & & & f_m^T \\\hline
\rowcolor{yellow!70} F_0 & G_1 & & & & f_1 & & & & & & & \\
\rowcolor{yellow!70} & F_1 & G_2 & & & & f_2 & & & & & & \\
\rowcolor{yellow!70} & & \ddots & \ddots & & & & \ddots & & & & & \\
\rowcolor{yellow!70} & & & F_{m-1} & G_m & & & &  f_m & & & & \end{array} \right) \left(\begin{array}{c} 
\rowcolor{red!30} v_0\\
\rowcolor{red!30} v_1 \\
\rowcolor{red!30} \vdots \\
\rowcolor{red!30} \vdots \\
\rowcolor{red!30} v_m \\\hline
 \eta_1 \\
 \eta_2 \\
 \vdots \\
 \eta_m \\\hline
\rowcolor{yellow!30} w_1 \\
\rowcolor{yellow!30} w_2 \\
\rowcolor{yellow!30} \vdots \\
\rowcolor{yellow!30}  w_m \end{array} \right) = -\left(\begin{array}{c} 
\rowcolor{red!70} 0 \\
\rowcolor{red!70} 0 \\
\rowcolor{red!70} \vdots \\
\rowcolor{red!70} \vdots \\
\rowcolor{red!70} 0 \\\hline
 0 \\
  \\
  \vdots\\
 0 \\\hline
\rowcolor{yellow!70} b_1\\
\rowcolor{yellow!70} b_2 \\
\rowcolor{yellow!70} \vdots \\
\rowcolor{yellow!70} b_m \end{array} \right)
\]}
\begin{gather*}
 F_i = \frac{I}{\Delta t} + \frac{1}{2}\frac{\partial f}{\partial u}(u_i,s), \quad G_i = -\frac{I}{\Delta t} + \frac{1}{2}\frac{\partial f}{\partial u}(u_i,s) ,\\ \quad b_i = \frac{1}{2} \left( \frac{\partial f}{\partial s}(u_i,s) + \frac{\partial f}{\partial s}(u_{i+1},s)\right), \quad f_i = \frac{1}{2} \left( f(u_i) + f(u_{i+1}) \right), \quad i = 0,...,m
\end{gather*}

This KKT system is a block matrix system, where each block ($I$, $G_i$, and $F_i$) is $n$ by $n$, 
where $n$ is the number of states (i.e. the product of the number of nodes and conservation variables in a compressible CFD 
simulation).  $w_i$ and $v_i$ are length $n$ vectors, and $\eta_i$ is a scalar.  The blocks highlighted in red correspond to equation (\ref{e:KKTw}), white to equation (\ref{e:KKTeta}), and yellow to (\ref{e:KKTv}).  

Note that as the system is symmetric, the adjoint is computed simply 
by changing the right hand side, allowing many gradients to be computed 
simultaneously \cite{Giles:2000:adj,Wang:2013:LSS2}.  

The KKT system is quite large, with $2mn + n + 1$ by $2mn + n + 1$ 
elements for $m$ time steps.  For a discretization with a stencil of five 
elements, the matrix would have approximately $23mn$ non-zero elements.  
Consider a 2D, compressible, computational fluid dynamics (CFD) simulation with $1\times10^5$ nodes ($n = 4\times10^5$) and 
16000 time steps. For this simulation, the KKT matrix would be $1.3\times10^{10}$ 
by $1.3\times10^{10}$ with $1.5\times10^{11}$ non-zero elements.  
The KKT system's large size and banded structure suggest that it should be solved with an iterative method. To this end, the authors of this paper have developed multigrid schemes that coarsen
the discretization of the system of interest in both space and time to solve the KKT system  \cite{Blonigan:2013:MG, gomez:2013:masters}.  Multigrid was chosen because equation \eqref{e:SPD_LSS} is a boundary value problem in time and because multigrid convergences very quickly relative to other iterative methods for many boundary value problems \cite{Briggs:2000:MGtutorial}. 

\subsection{LSS type II for Homogeneous Isotropic Turbulence} \label{sec_lss:isoturb}

The first fluid dynamics problem LSS has been demonstrated on is the sensitivity of the cumulative energy spectrum of a fluid flow with Homogeneous Isotropic Turbulence (HIT). A traditional HIT solver involves a direct numerical simulation of the three-dimensional Navier-Stokes equations on a cube of fluid. The boundaries of the cube are periodic in each direction, and the fluid is forced, deterministically, to drive the flow \cite{isoturb_2009}. Given the correct choice of forcing $F$, the result is a fully three-dimensional, unsteady, flow, that is chaotic with some bounded energy distribution. Typically, this flow is assumed to be incompressible, as shown in equation \eqref{eq_lss:iso_space}, where $U$ is the velocity field and $P$ is the pressure field.

\begin{align} \label{eq_lss:iso_space}
  \pdt{}U(x,t) + \nabla\cdot\bigg(U(x,t)\otimes U(x,t)\bigg) &= -\nabla P(x,t) + \nu\nabla^2 U(x,t) + F \left(\ U(x,t)\ \right) \\
  \nabla \cdot U(x,t) &= 0 \nonumber
\end{align}
The simple, periodic, domain lends itself to a pseudo-spectral solution method \cite{Orszag:1972:turb}. Given a spatial Fourier transform $\mathcal{F}$, equation \eqref{eq_lss:iso_spec} shows the pseudo-spectral transformation of incompressible Navier-Stokes. $\hat{G}$ represents the nonlinear convection operator, and is computed in the physical domain \eqref{eq_lss:iso_nonlin}.
\begin{equation}
  \Uh = \mathcal{F}\left[U(x,t) \right]
\end{equation}
\begin{align} \label{eq_lss:iso_spec}
  \pdt{}\Uh + \hat{G}\left( \Uh \right) &= -i \mathbf{k} \Ph - \nu |\mathbf{k}|^2 \Uh + \hat{F}\left(\ \Uh \right) \\
  \mathbf{k}\cdot\Uh &= 0 \nonumber
\end{align}
\begin{align} \label{eq_lss:iso_nonlin}
	U(x,t) &= \mathcal{F}^{-1}\left[\Uh \right] \nonumber \\
  \hat{G}\left( \Uh \right) &= i\mathbf{k}\cdot\mathcal{F}\left[\ U\otimes U \ \right]
\end{align}

The spectral forcing, $\hat{F}(\hat{U})$, is chosen to inject energy at some rate to balance the energy loss from viscosity. In the spectral domain this forcing is applied to the lower frequency modes ($\mathbf{\tilde{k}}$) with wavenumber given in \eqref{eq_lss:ktilde}.
\begin{equation} \label{eq_lss:ktilde}
	\mathbf{\tilde{k}}_0 \equiv (1,0,0),\quad \mathbf{\tilde{k}}_1 \equiv (0,1,0),\quad \mathbf{\tilde{k}}_2 \equiv (0,0,1)
\end{equation}
The energy $\varepsilon$ is computed on these low-frequency modes \eqref{eq_lss:lowE}. The forcing is proportional to some constant power $\mathbb{P}$ divided by the low-frequency energy $\varepsilon$, and proportional to the low-frequency spectral velocity \eqref{eq_lss:isoForce}. The higher-frequencies are left unforced.
\begin{align}
	&\varepsilon = \sum_{i = 0}^{2} \left| \hat{U}( \mathbf{\tilde{k}}_i, t ) \right|^2 \label{eq_lss:lowE}\\
	&\hat{F}(\mathbf{\tilde{k}}_i,\ t) = \frac{\mathbb{P}}{\varepsilon}\ \hat{U}(\mathbf{\tilde{k}}_i,\ t) \label{eq_lss:isoForce} \\
	&\hat{F}(\mathbf{k}\not\in \mathbf{\tilde{k}},\ t) = 0 \nonumber
\end{align}
The spectral forcing coefficient, $\mathbb{P}$, may be chosen to achieve the desired Taylor micro-scale Reynolds number.  $\beta$ is a fractional deviation from the nominal power coefficient, and will be the perturbation parameter considered here, as shown in equation \eqref{eq_lss:pow_coeff}.
\begin{equation} \label{eq_lss:pow_coeff}
	\mathbb{P} \equiv \mathbb{P}_0\ (1+\beta)
\end{equation}

The objective function considered in this problem is the cumulative energy spectrum of the flow field \eqref{eq_lss:cumspec}, and its long-time average \eqref{eq_lss:spec_mean}. When increasing $\beta$, the energy at all values of the wavenumber magnitude ($|\mathbf{k}|$) are expected to increase on average, however, the exact amount is difficult to obtain. One primal flow field will be simulated and that flow field, or trajectory, is then used to predict the sensitivity of the cumulative energy spectrum $d \mean{E}(|\mathbf{k}|)/d\beta$, across all frequencies.
\begin{align} \label{eq_lss:cumspec}
	E(|\mathbf{k}|,\ t) = \sum_{k \geq |\mathbf{k}|} \left|\hat{U}(k,\ t)\right|^2 \\
	\mean{E}(|\mathbf{k}|) = \lim_{T\rightarrow\infty}\ \frac{1}{T} \int_0^T E(|\mathbf{k}|,\ t) \ dt \label{eq_lss:spec_mean}
\end{align}

\noindent In the notation of section \ref{c:LSS}:

\[
u \equiv \hat{U} \qquad \mean{J} \equiv \mean{E}(|\mathbf{k}|) \qquad s \equiv \beta
\]

\noindent In this case $f(u)$ refers to the sum of all operators excluding $\pd{}{t}$ in equation (\ref{eq_lss:iso_spec}).

\subsubsection{Simulation Details}

The spatial domain is a cube of size $[0,\; 2\pi]^3$, each velocity component is discretized on a ($3k\times 3k\times 3k$) uniform Cartesian grid ($U \in \mathbb{R}^{3\times 3k\times 3k\times 3k}$). In the spectral domain the $2/3$ rule is applied to avoid aliasing, 
giving a maximum wave number of $2k$ in each dimension. Additionally because the flow is real valued, the Fourier modes will be symmetric and may be further reduced by about a half ($\hat{U} \in \mathbb{C}^{ 3\times 2k\times 2k\times (k+1) }$). The discrete Fourier transforms, and inverse transforms are computed via a real valued Discrete Fourier Transform \cite{anfft,fftw_1}.

For the purposes of the LSS solver, and to emphasize the generality of the method, these equations will be treated as a black-box differential equation. The discrete isotropic flow solver will implement a simple interface, providing access to evaluating $f$ and multiplication by $\pdu{f}$ and $\left(\pdu{f}\right)^*$. In fact, the LSS solver does not even know that the isotropic turbulence model is solved in the complex domain, it simply treats the discretized solution as real system of twice the size, ($M_k = 2\times 3\times 2k\times 2k \times (k+1) = 24k^2(k+1)$). 



\subsubsection{Results}
\begin{figure}[htp!]
	\centering
	\includegraphics[width=\textwidth]{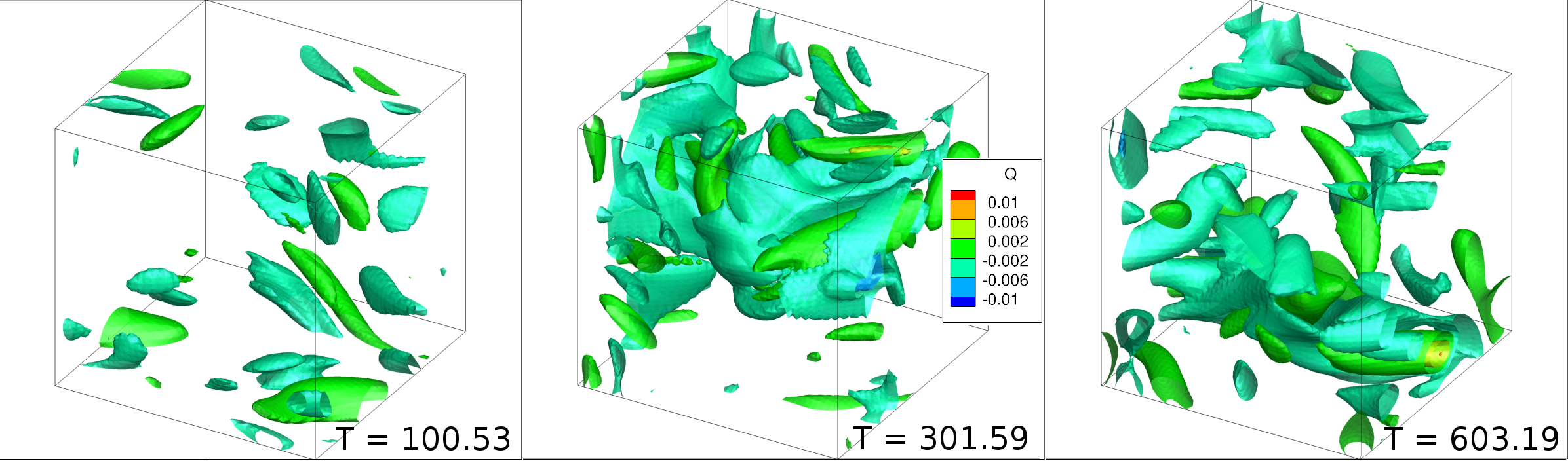}
	\caption{Q-Criterion iso-surfaces for several primal ($U$) solutions ($Re_\lambda = 33.06$).  Note that $Q = \hf\left(\|\Omega\|_f^2 - \|S\|_f^2\right)$, $\Omega = \hf\left(\nabla U - \nabla U^T\right)$, $S = \hf\left(\nabla U + \nabla U^T\right)$\cite{Hunt:1988:Eddy}.    }
	\label{fig_iso:primal}
\end{figure}

\begin{figure}[htp!]
	\centering
	\includegraphics[width=.6\textwidth]{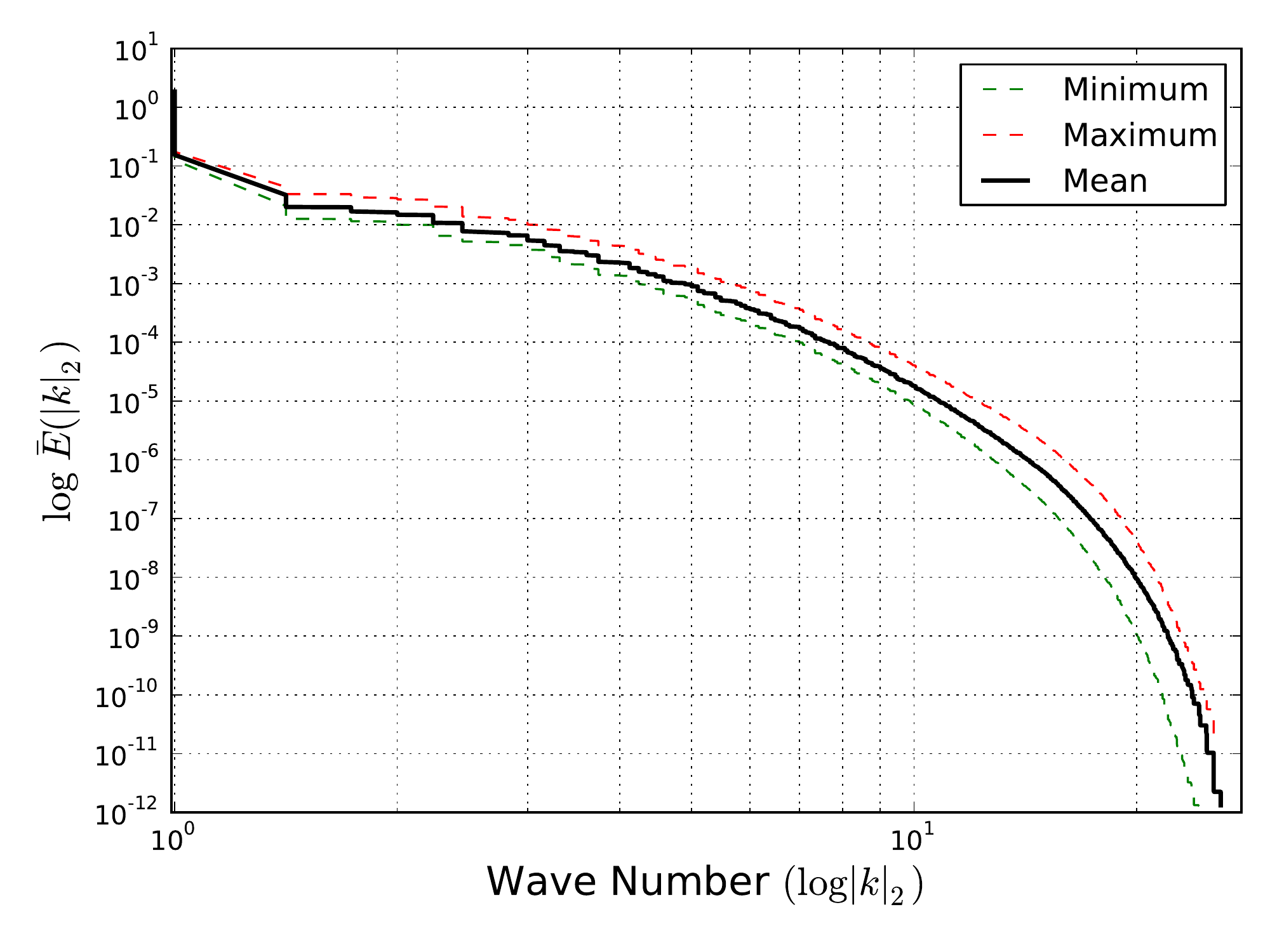}
	\caption{Average cumulative energy spectrum $\mean{E}$, for primal solution to HIT}
	\label{fig_iso:primal_spec}
\end{figure}

One primal solution of the Isotropic Flow solver is produced using the discretized isotropic flow equations. The solution is obtained using the same implicit trapezoidal method described in section \ref{s:typeII}. For all future results assume the isotropic flow is solved using the following parameters.
\begin{align*}
	\nu &= 0.01 \\
	k &= 16 \\
	\Delta t &= \frac{2\pi}{16} \\
        \mathbb{P}_0 &= 0.02 
\end{align*}
The Taylor micro-scale for this flow is $\lambda = 1.22$ and the associated Reynolds number $Re_\lambda = 33.06$. The primal solution is solved for $N+1$ time-steps, each of size $\Delta t = \frac{2\pi}{16}$. Here, $N$ will be assumed to be $2^{12} = 4096$, giving a total time domain of size $T=1608.50$. Several example primal solutions are shown in figure \ref{fig_iso:primal}. For reference this time domain size is approximately $6.96$ times the length of the longest time scale ($t_\text{ref} = 2\pi/\bar{U}$). Figure \ref{fig_iso:primal_spec} shows the mean energy spectrum for one such run. The LSS equations for the long scale $N=2^{12}$ problem are solved utilizing $n_p = 64$ processes. Because of a limitation in the implementation of the parallelization, the multigrid cannot coarsen in time below $n_p$ time-steps. The time-dilation weighting, defined in section \ref{s:typeII}, $\alpha = 10\sqrt{10}$, is used. 

\begin{figure}
	\centering
	\includegraphics[width=0.6\textwidth]{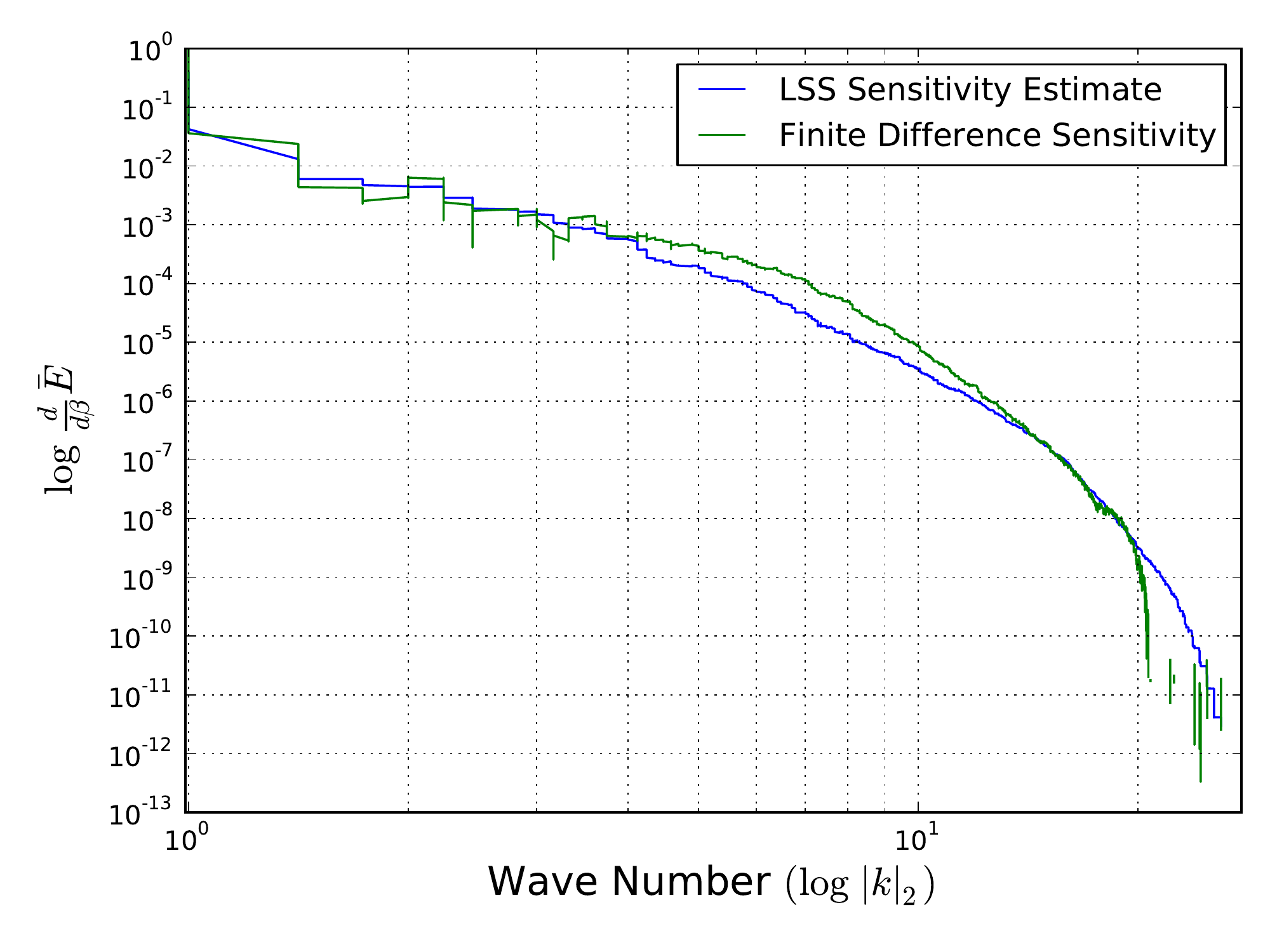}
	\caption{LSS vs. Finite Difference sensitivity estimates. Finite differences computed using two primal solutions with the forcing magnitude altered by $\pm 5 \% $ ($\Delta\beta = \pm 0.05$). LSS estimate computed using parallel multigrid and equation \eqref{e:LSSgrad}. }
	\label{fig_iso:compare}
\end{figure}

The climate sensitivity is then computed in two ways. A finite difference estimate of the $\beta$ sensitivity is performed. $\beta$ is perturbed by some $\Delta\beta = 0.05$ which produces a $5\%$ change to the input power $\mathbb{P}$. 
\begin{equation}
	\beta^{+} = 0.05 \qquad \beta^{-} = -0.05
\end{equation}
Two new primal trajectories are simulated with these two perturbed $\beta$ values. 
\begin{align}
	\hat{U}^{+} &= \hat{U}(\mathbf{k},\ t;\ \beta^{+}) \\
	\hat{U}^{-} &= \hat{U}(\mathbf{k},\ t;\ \beta^{-})
\end{align}
The difference in the instantaneous cumulative energy spectrum \eqref{eq_lss:spec_mean} between the two perturbed runs is computed, and time-averaged. 
\begin{align}
	E^{+} = E(|k|, t)\bigg|_{\hat{U}^+} &\qquad E^{-} = E(|k|, t)\bigg|_{\hat{U}^-} \\
	\dd{}{\beta}E(|k|,\ t) &\approx \frac{ E^{+} - E^{-} }{2\Delta\beta} \\
	\dd{}{\beta}\mean{E}(|k|)_{(FD)} &= \frac{1}{N+1} \sum_{i = 0}^N \frac{ E_i^{+} - E_i^{-} }{2\Delta\beta}
\end{align}
These are compared against the sensitivity estimate given by the LSS equations given by the unperturbed trajectory ($U$). The LSS solution is obtained using a parallel multigrid algorithm to compute $w$, $v$, and $\eta$ \cite{gomez:2013:masters}.
Then the sensitivities from that solution are approximated using equation \eqref{e:LSSgrad}.
\begin{align}
	\mean{\eta} &= \frac{1}{N} \sum_{j=0}^{N-1} \eta_j \\
	\hat{U}^m_j &= \hf\lpar{\hat{U}_j + \hat{U}_{j+1}} \\ 
	\pd{}{\beta}\mean{E}(|k|)_{(LSS)} &= \frac{1}{N+1}\sum_{i=0}^N \pd{E}{\hat{U}}\bigg|_{\hat{U}_i} v_i
			+ \frac{1}{N}\sum_{j=0}^{N-1}\ (\mean{\eta} - \eta_j)\ E\bigg|_{\hat{U}^m_j}
\end{align}
Figure \ref{fig_iso:compare} shows how these two methods compare. As observed for smaller ODE systems \cite{Wang:2013:LSS2} and other PDEs \cite{Blonigan:2013:LSS,Blonigan:2013:KS}, the LSS method produces fairly good sensitivity estimates compared to finite difference. 


\section{LSS type III: Checkpoint Design}
\label{s:LSSIII}

\newcommand{\hv}{\hat{v}}
\newcommand{\hw}{\hat{w}}
\newcommand{\bv}{\textbf{v}}
\newcommand{\bw}{\textbf{w}}
\newcommand{\hbv}{\hat{\textbf{v}}}
\newcommand{\hbw}{\hat{\textbf{w}}}
\newcommand{\prop}[2]{\phi^{{#1},{#2}}}

Although LSS type II has shown promise, the large size of the KKT system makes the application of type II very computationally expensive and time consuming.  Using multigrid in space and time, solving the KKT system for the homogeneous isotropic turbulence simulation discussed in the previous section took days.  

\begin{figure}
\centering
\includegraphics[width = 0.7\textwidth]{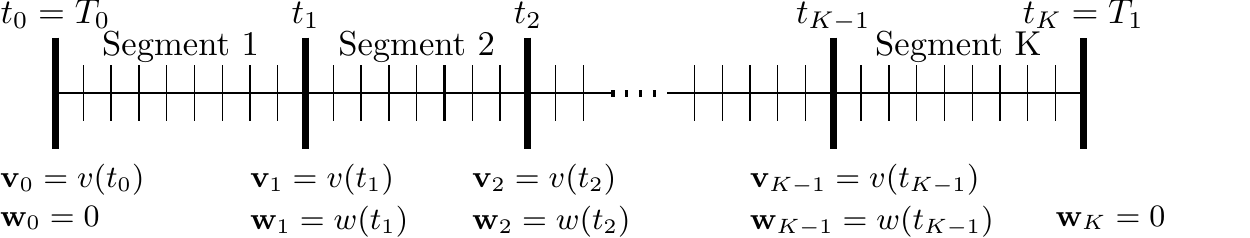}
\caption{Diagram showing the checkpoints and time segments used in LSS type III.  }
\label{f:time_segments}
\end{figure}

LSS type III offers a means to reduce the size of the KKT system.  Instead of searching for the entire shadow trajectory $u(t)$ at once, LSS type III searches for the shadow trajectory at $K+1$ checkpoints, $t_0,t_1,t_2,...,t_K$, shown in the diagram in figure \ref{f:time_segments}.  

The spacing between checkpoints is chosen to ensure that $v(t)$ does not grow too much between $t_i$ and $t_{i+1}$ or $w(t)$ does not grow too much between $t_{i+1}$ and $t_i$.  

To explain how LSS type III works, we first consider a case with one time segment.  Note that this is equivalent to LSS type I. Assuming we have already obtained the primal solution $u(t)$, the tangent is solved as follows:  

\begin{enumerate}
\item Solve tangent equation \eqref{e:KKTv} from some initial condition $v(T_0) = \bv_0$.  
\item Solve the Lagrange multiplier equation \eqref{e:KKTw} backward in time with $w(T_1) = 0$.  
\item Solve for residual ${\bf R}^v_0 = w(T_0) - 0$.  Use ${\bf R}^v_0$ to update our initial condition $\bv_0$ and repeat steps 1-3 until convergence.  
\end{enumerate}

As stated in section \ref{s:LSform}, the condition number of LSS type I is of the order $e^{\Lambda T}$, for largest positive Lypunov exponent $\Lambda$ and simulation time $T$. In other words, the condition number scales with a normalized perturbation to $v(t)$, $\delta v$:

\[
\kappa_I \sim \frac{\|\delta v\|e^{\Lambda T}}{\|\delta v\|} = e^{\Lambda T}
\]

Therefore, we can reduce the condition number of this problem by guessing what $v(t)$ is at the checkpoint times $t_0,t_1,...,t_{K-1}$.  This allows perturbations to the tangent equation to grow only within each time segment.  Therefore, the condition number for LSS type III is of the order $e^{\Lambda\Delta T_{max}}$, where $\Delta T_{max}$ is the size of the large time segment.  An outline of the algorithm with multiple time segments is:

\begin{enumerate}
\item Solve tangent equation \eqref{e:KKTv} in all $K$ time segments from some initial condition $v(t_i) = \bv_i, \quad i = 0,1,...,K-1$.  
\item Solve the lagrange multiplier equation \eqref{e:KKTw} backward in time in each time segment.  For the last time segment, solve from the terminal condition $w(t_K) = 0$, otherwise solve from some terminal condition $w(t_i) = \bw_i, \quad i = 1,...,K-1$ .  
\item Solve for the residuals ${\bf R}^v_0 = w(t^+_0) - 0$, ${\bf R}^v_i = w(t^+_i) - \bw_i, i = 1,...,K-1$ and ${\bf R}^w_i = v(t^-_i)-\bv_i, i = 1,...,K-1$.  Use ${\bf R}^v_i$ to update each $\bv_i$ and ${\bf R}^w_i$ to update each $\bw_i$.  Repeat steps 1-3 until convergence.  
\end{enumerate}

\noindent This algorithm can be thought of as an iterative solver of the following system of linear equations:  

\begin{equation}
\uuline{A}\uline{x} = \uline{b}, \qquad \uline{x} \equiv \left(\begin{array}{c}
\uline{\bv} \\
\uline{\bw}
\end{array} \right)
\label{e:tanLSSsys}
\end{equation}

\noindent where $\uline{\bv}$ and $ \uline{\bw}$ represent vectors of initial conditions $\bv_i$ and terminal conditions $\bw_i$. From the algorithm outline, we see that the $\uline{\bv}$ is a length $nK$, and $\uline{\bw}$ is a length $n(K-1)$, where $n$ is the number of states of the system.  Therefore, $\uuline{A}$ can be thought of as a $n(2K-1)$ by $n(2K-1)$ system.  For more details on the structure of $\uuline{A}$, refer to appendix \ref{s:tanmat}.  

Iterative solvers such as MINRES require the right hand side $\uline{b}$ and the matrix multiplication operation $\uuline{A}\uline{x}$. But before we can form algorithms for these operations, a few minor details need to be addressed.  In order to solve the tangent equation \eqref{e:KKTv} and compute gradients with \eqref{e:LSSgrad}, we need a way to compute the contribution of $\eta$.  It can be shown (See Appendix \ref{s:tanform}) that

\begin{equation}
\langle f, v \rangle = 0
\label{e:v_orth}
\end{equation}

\noindent and that if we solve the equation $\frac{dv'}{dt} = \pd{f}{u} v' + \pd{f}{s}$, then:

\begin{equation}
v(t) = P_t v'(t) \equiv v'(t) - \frac{v'(t)^T f(t)}{f(t)^T f(t)} f(t)
\label{e:v_proj}
\end{equation}

Additionally, to enforce \eqref{e:KKTeta} and \eqref{e:v_orth}, the projection operator $P_t$ defined in equation \eqref{e:v_proj} is applied to $v(t_i^-)$ and $w(t_i^+)$.  Additionally, the identities 
$\bv_i = P_{t_i}\bv_i$ and $\bw_i = P_{t_i}\bw_i$ are used.  This makes the linear operator $\uuline{A}$ associated with the LSS type III algorithm symmetric, as shown in appendix \ref{s:tanmat}.  

Finally, it is important that a dual consistent time-stepping scheme is used to solve the tangent and Lagrange multiplier equations \eqref{e:KKTv} and \eqref{e:KKTw}.  This ensures that the linear operator in equation \eqref{e:tanLSSsys} is symmetric. Altogether: 

\begin{framed}
\noindent\textbf{Tangent MATVEC Algorithm:}  
Inputs: $\bv_i$, $\bw_i$, $\beta$; 
Ouputs: ${\bf R}^w_i$, ${\bf R}^v_i$ \\
NOTE: In terms of equation \eqref{e:tanLSSsys}, MATVEC computes $\uline{r} = \uuline{A}\uline{x} - \beta\uline{b}$
\begin{enumerate}
\item In each time segment, time integrate
$\frac{dv'}{dt} = \pd{f}{u} v' + \beta\pd{f}{s},\; t\in(t_i,t_{i+1}]$
with initial conditions $v'(t_i) = P_{t_i}{\bf v}_i$. 
\item In each time segment, time integrate
$\frac{dw}{dt} = -\left(\pd{f}{u}\right)^* w - P_{t}\,v'(t) ,\;
t\in(t_i,t_{i+1})$ with terminal condition $w(t_{i+1}) = P_{t_{i+1}}{\bf w}_{i+1}$.  
\item Compute ${\bf R}^w_i = {\bf v}_i - P_{t_i}\,v'(t_i^-)$.  
\item Compute ${\bf R}^v_i = {\bf w}_i - P_{t_i}\,w(t_i^+)$.
\end{enumerate}

\noindent\textbf{Tangent LSS type III Solver} 
\begin{enumerate}
\item Start with an initial guess for ${\bf v}_i$, ${\bf w}_i$.  
\item To form the right hand side of the linear system, $\uline{b}$, use MATVEC algorithm with $\beta = -1$ and ${\bf v}_i$, ${\bf w}_i$ = 0. 
\item Use iterative algorithm of choice.  To compute $\uuline{A}\uline{x}$, use MATVEC with $\beta = 0$.  
\item To compute the gradient $d\bar{J}/ds$ use MATVEC with $\beta = 1$ to find $v'(t)$, then solve:
\begin{equation}
\frac{d\bar{J}}{ds} =  
\frac1T \sum_{i=0}^{K-1}
 \int_{t_i}^{t_{i+1}} \frac{\partial J}{\partial u}\bigg|_{t}^T v'  \ dt
+ \frac1T \sum_{i=0}^{K-1} \frac{f_{i+1}^T v'(t_{i+1})}{f_{i+1}^T f_{i+1}} \left(\bar{J} - J(t_{i+1})\right)
\label{e:LSSIIIgrad}
\end{equation} 
where $f_{i+1} = f(u(t_{i+1}))$ and $T \equiv T_1-T_0 = t_K-t_0$.  The derivation of equation \eqref{e:LSSIIIgrad} is in appendix \ref{s:tanform}. 
\end{enumerate}
\end{framed}

From tangent LSS type III, we can derive adjoint LSS type III, so that we can efficiently compute the sensitivity of one time averaged objective function $\bar{J}$ to many design parameters $s$.  This is done by finding the discrete adjoint of equation \eqref{e:tanLSSsys}, as shown in appendix \ref{s:adjform}.  This results in an adjoint algorithm similar to the tangent algorithm, where $\hv$ and $\hw$ are the adjoint variables:

\begin{framed}
\noindent\textbf{Adjoint MATVEC Algorithm:}   
Inputs: $\hbv_i$, $\hbw_i$, $\beta$; 
Ouputs: ${\bf R}^{\hw}_i$, ${\bf R}^{\hv}_i$ \\
NOTE: Expressing adjoint LSS type III as a linear system, MATVEC computes $\uline{r} = \uuline{A}\uline{x} - \beta\uline{b}$, where $\uuline{A}$, $\uline{x}$, and $\uline{b}$ are defined for the adjoint formulation in appendix \ref{s:adjform}.  
\begin{enumerate}
\item In each time segment, time integrate
$\frac{d\hv'}{dt} = \frac{\partial f}{\partial u} \hv',\; t\in(t_i,t_{i+1}]$
with initial conditions $\hv'(t_i) = P_{t_i}{\hbv}_i$. 
\item In each time chunk, time integrate
$\frac{d\hw}{dt} = -\left(\frac{\partial f}{\partial u}\right)^* \hw - P_{t}\,\hv'(t) + \beta \frac1T\pd{J}{u} ,\;
t\in(t_i,t_{i+1})$ with terminal condition $\hw(t_{i+1}) = P_{t_{i+1}}{\hbw}_{i+1} - \beta\frac{J(t_{i+1}) - \bar{J}}{T f_{i+1}^Tf_{i+1}} f_{i+1}$.  
\item Compute ${\bf R}^{\hw}_i = {\hbv}_i - P_{t_i}\,\hv'(t_i^-)$.  
\item Compute ${\bf R}^{\hv}_i = {\hbw}_i - P_{t_i}\,\hw(t_i^+)$. 
\end{enumerate}

\noindent\textbf{Adjoint LSS type III Solver} 
\begin{enumerate}
\item Start with an initial guess for ${\hbv}_i$, ${\hbw}_i$.  
\item To form right hand side of the linear system, use MATVEC algorithm with $\beta = -1$ and ${\hbv}_i$, ${\hbw}_i$ = 0. 
\item Use iterative algorithm of choice.  To conduct matrix multiplication, use MATVEC with $\beta = 0$.  
\item To compute the gradients $d\bar{J}/ds$, use MATVEC with $\beta = 1$ to find $\hw(t)$ then solve: 
\begin{equation}
 \dd{\bar{J}}{s} = \sum_{i=0}^{K-1} \int_{t_i}^{t_{i+1}} \pd{f}{s}\bigg|_{t}^T \hw(t) \ dt 
\label{e:AdjGrad}
\end{equation}
\end{enumerate}
\end{framed}

\noindent By inspecting the tangent and adjoint algorithms, we see that an LSS type III solver can be assembled from four components:

\begin{enumerate}
\item A primal solver, which finds $u(t)$ by solving the primal equation $\dd{u}{t} = f(u;s)$.  
\item A tangent solver, which solves the linearized primal or tangent equation 
\[\dd{v'}{t} = \pd{f}{u}v' + F\]
where $F=\pd{f}{s}$ for tangent LSS and $F=0$ for adjoint LSS.  
\item An adjoint solver, which solves the Lagrange multiplier equation 
\[\dd{w}{t} = -\left(\pd{f}{u}\right)^*w - v + F\]
where $F = 0$ for tangent LSS and $F= \frac1T\pd{J}{u}$ for adjoint LSS.  
\item A projection operator $P_t$ as defined in equation \eqref{e:v_proj}.  
\end{enumerate}

There are well established methods to construct each of these four components.  Therefore, LSS type III can be implemented as a framework of existing solvers.  

\subsection{LSS type III for the Kuramoto-Sivashinsky equation}

The Kuramoto-Sivashinsky (KS) equation is a 4th order, chaotic PDE, that can be used to model a number of physical phenomena including turbulence in the Belouzov-Zabotinskii reaction and thermal diffusive instabilities in laminar flame fronts \cite{Hyman:1986:KS}:
\[ 
\frac{\partial u}{\partial t} = 
-(u + c) \frac{\partial u}{\partial x}
- \frac{\partial^2 u}{\partial x^2}
- \frac{\partial^4 u}{\partial x^4}
\]

\noindent The $c$ term is added to make the system ergodic \cite{Blonigan:2013:LSS}.  In addition, the KS equation is made ergodic by using the boundary conditions: 
\[ u\Big|_{x=0,128} = \frac{\partial u}{\partial x}\bigg|_{x=0,128}
 = 0 \]
The objective function $\bar{J}$ was chosen to be $u$ averaged over both space and time:
\[
\bar{J} = \frac{1}{128T} \int_0^T \int_0^{128} u(x,t) \ dx \ dt
\]

\noindent First, we demonstrate LSS type III on the KS equation with $c = 0.5$ and the following initial condition:

\[
 u(x,0) = \bigg\{\begin{array}{c}
           1, \quad x = 64 \\
           0, \quad x \neq 64
          \end{array}
\]

\begin{figure}
\centering
\includegraphics[width = \textwidth]{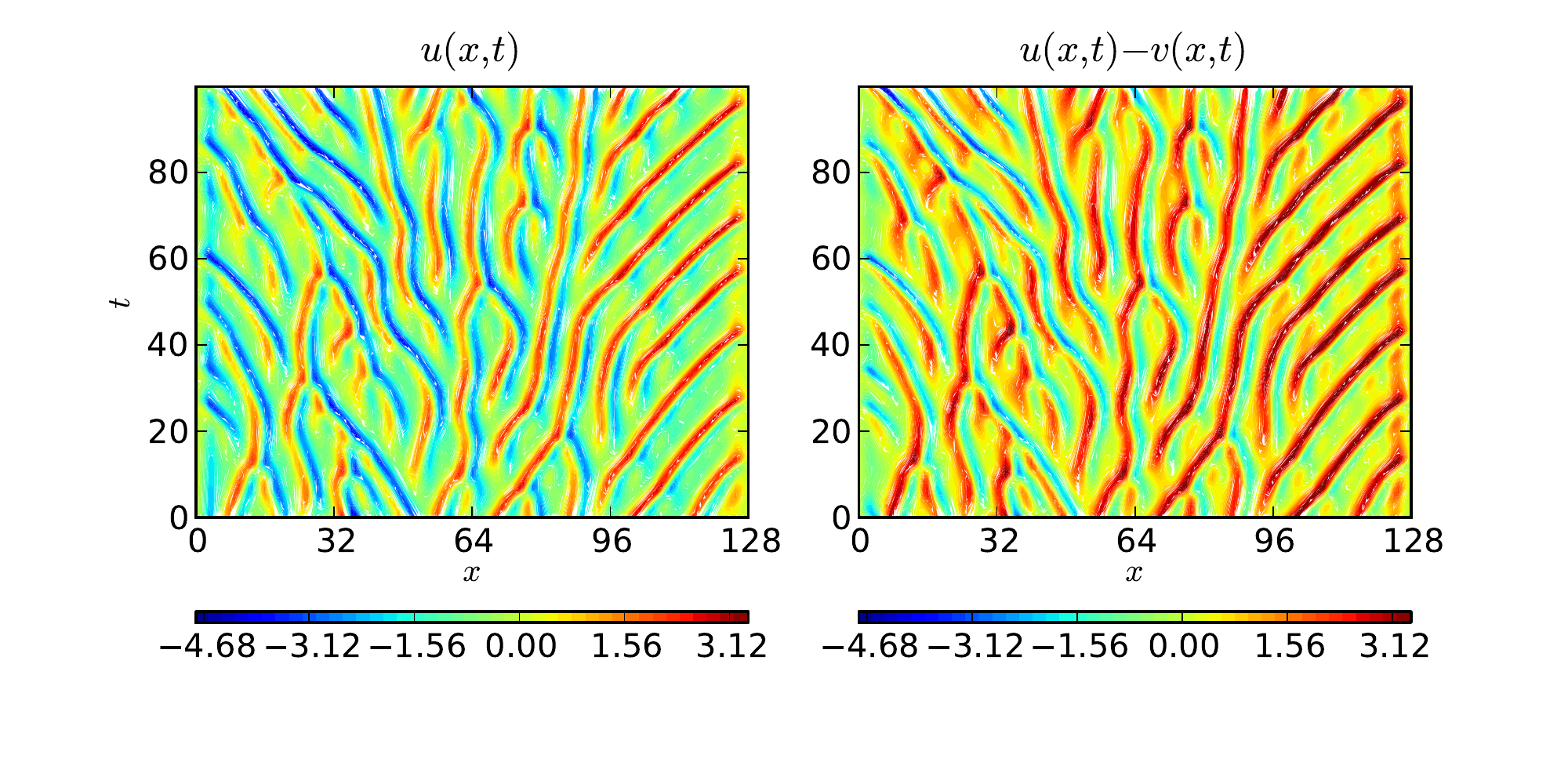}
\caption{LEFT: Primal solution $u(x,t)$ of the K-S equation for $c=0.5$.  RIGHT: First order approximation $u(x,t) - v(x,t)$ of the shadow trajectory for $c=-0.5$.  }
\label{f:KSprimal}
\end{figure}

\noindent The KS equation was discretized using a 2nd order finite difference scheme in space and solved forward in time using a 3rd order explicit, dual consistent, Runge-Kutta time stepping scheme\cite{Yang:2011}.  The tangent and adjoint solvers used the same Runge-Kutta scheme.  A node space of $\Delta x = 1.0$ and a time step size of $\Delta t = 0.2$ was used.  The system was integrated for 500 time units before LSS was applied to ensure the solution was at a quasi-steady state (on the attractor).  After this run-up, the primal was run for $T=100$ time units.  This primal solution $u(x,t)$ is shown in figure \ref{f:KSprimal}.  

\begin{figure}
\centering
\includegraphics[width = \textwidth]{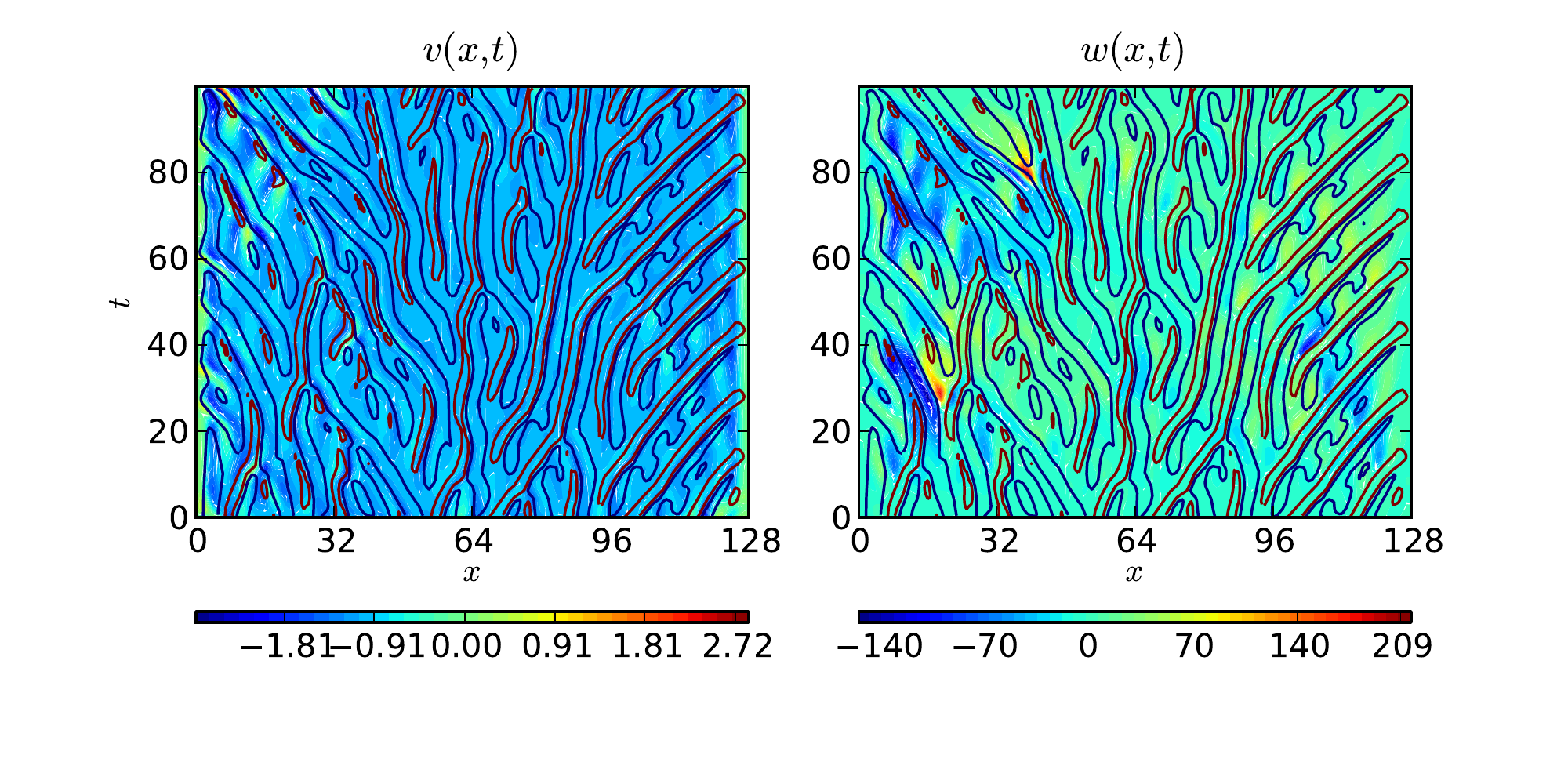}
\caption{LEFT: tangent solution $v(x,t)$ of the K-S equation for $c=0.5$.  RIGHT: Corresponding Lagrange multipliers $w(x,t)$. Contour lines corresponding to $\pm30\%$ of $|u(x,t)|$ are overlayed on both plots.  }
\label{f:KStan}
\end{figure}

LSS type III was applied with $25$ time segments of length $\Delta T =4$.  The MINRES method was used to implement both tangent and adjoint LSS type III.  The algorithm was run until the relative residual L2 norm was less than $10^{-6}$.  $x-t$ diagrams of the solutions $v(x,t)$ and $w(x,t)$ for tangent LSS type III are shown in figure \ref{f:KStan}.  Additionally, a low order approximation to a shadow trajectory is shown in figure \ref{f:KSprimal}.  The approximate shadow trajectory has similar spatial-temporal turbulent structures to the primal $u(x,t)$, but with larger values of $u(x,t)$.  The predicted gradient was $\dd{\bar{J}}{c} = -0.9597$, which is similar to the sensitivity values computed using LSS type II with simulations times of $100$ time units observed in the author's previous work \cite{Blonigan:2013:KS}.   

\begin{figure}
\centering
\includegraphics[width = \textwidth]{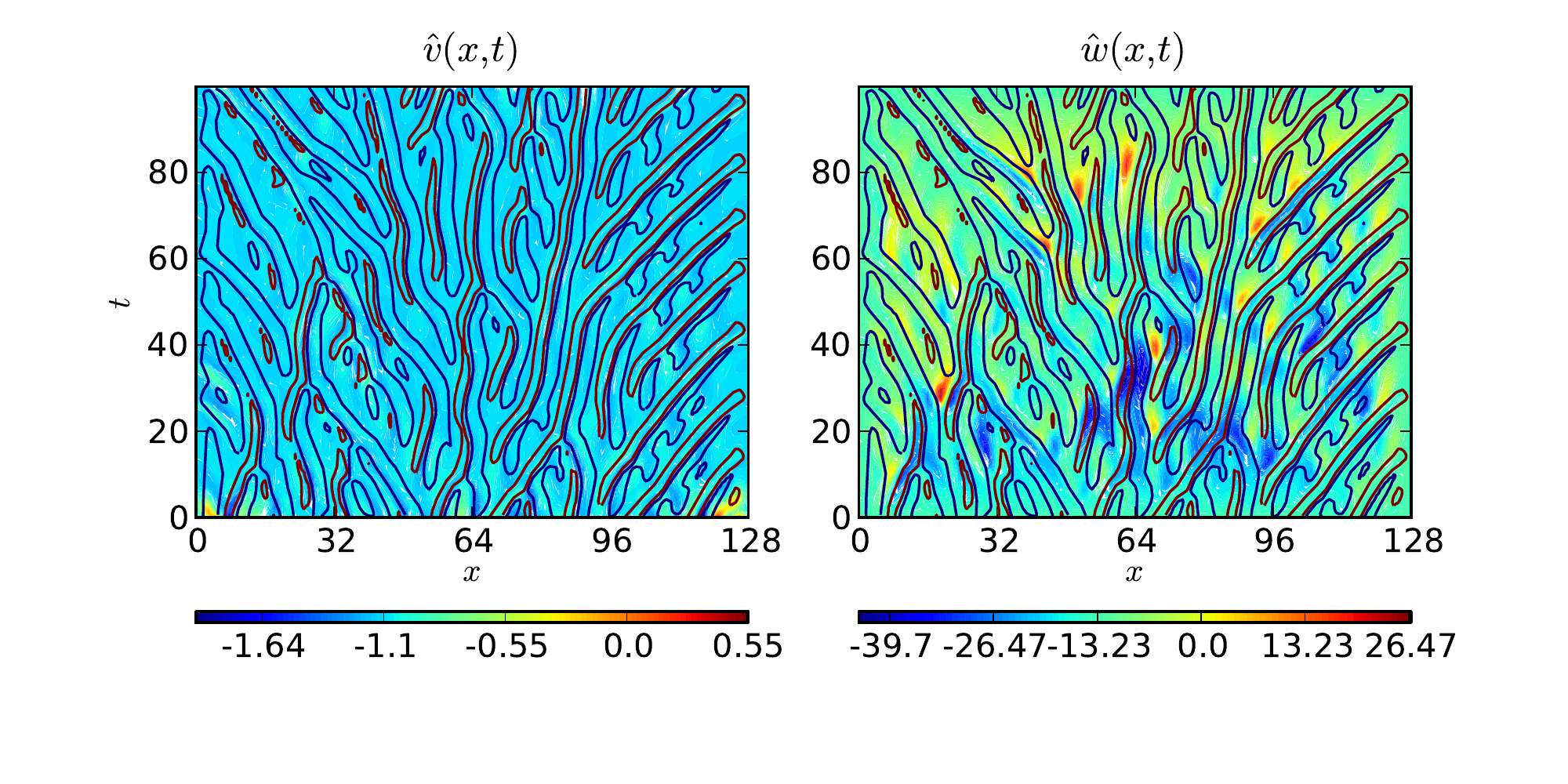}
\caption{LEFT: adjoint Lagrange multipliers $\hv(x,t)$ of the K-S equation for $c=0.5$.  RIGHT: Corresponding adjoint solution $\hw(x,t)$. Contour lines corresponding to $\pm30\%$ of $|u(x,t)|$ are overlayed on both plots.  Note that $\hv(x,t)$ and $\hw(x,t)$ are scaled by $10^4$ in the contour plots.  }
\label{f:KSadj}
\end{figure}

The same primal $u(x,t)$ solution was analyzed with adjoint LSS type III.  $x-t$ plots of the adjoint solutions $\hv(x,t)$ and $\hw(x,t)$ are shown in figure \ref{f:KSadj}. Adjoint LSS computed a gradient of $\dd{\bar{J}}{c} = -0.9587$, very similar to the result from tangent LSS.  This slight difference arises from how the forcing terms ($\pd{f}{s}$ in tangent equation, $v + \frac1T\pd{J}{u}$ in the Lagrange multiplier equation) are discretized.  In the tangent solver, forcing is applied for every sub-step of the Runge-Kutta scheme.  In the adjoint solver used to solve the Lagrange multiplier equations, forcing was applied before the first sub-step of the Runge-Kutta scheme.

\begin{figure}
\centering
\includegraphics[width = 0.5\textwidth]{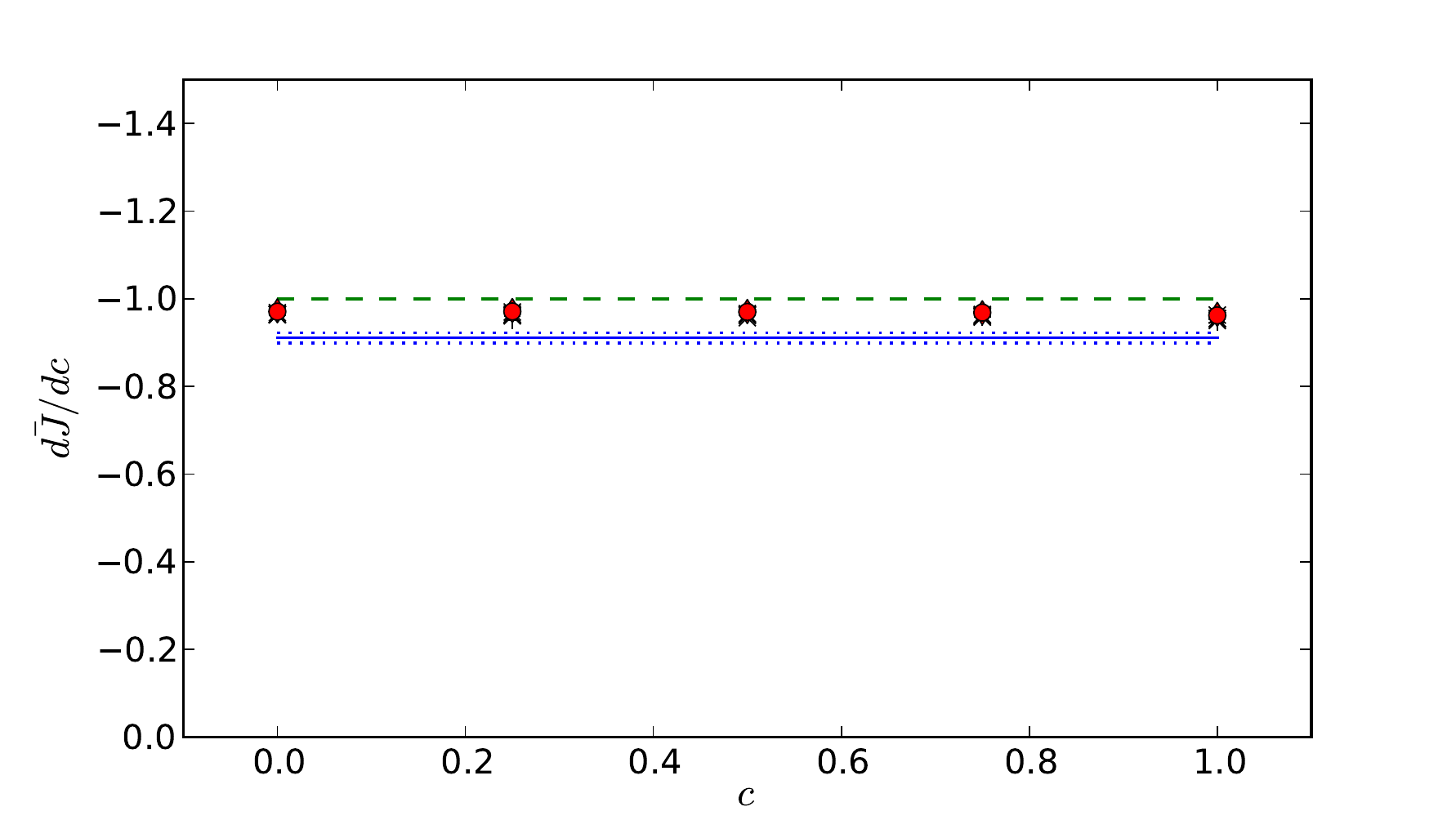}
\caption{Gradients computed using tangent and adjoint LSS type III compared with linear regression data (in blue, with $\pm3\sigma$ as dotted lines) and LSS type II solutions for $T=1000$ (green dotted line).  Tangent and adjoint gradients computed with $T=100$ are indicated by black o's and x's, respectively.  Tangent and adjoint gradients computed with $T=1000$ are indicated by red diamonds and circles, respectively.  
LSS type II and linear regression data was obtained from Blonigan and Wang\cite{Blonigan:2013:KS}.  Note that $\sigma$ is the standard error of the linear regression\cite{Blonigan:2013:KS}.  }
\label{f:KSgrad}
\end{figure}

To validate our results, we compare gradients computed using tangent and adjoint LSS type III to gradients computed using LSS type II and linear regression. A MINRES solver was used on a $T=100$ simulation with $25$ time segments of length $\Delta T =4$ and a $T=1000$ simulation with $250$ time segments of length $\Delta T =4$.  Gradients were found for five values of $c$, with the initial condition $u(x,0)$ chosen randomly from $[-0.5,0.5]$ for all $x$.  We computed gradients from five different initial conditions for $T=100$ and one initial condition for $T=1000$ for each value of $c$.  In figure \ref{f:KSgrad}, we see that the gradients computed using tangent and adjoint LSS type III are virtually indistinguishable from one another.  Like LSS type II, type III slightly overestimates the gradients relative to the linear regression results, but this overestimation is smaller.  The reason for the small discrepancy between LSS and the linear regression is still unclear to the authors at this time.  

Finally, it is important to compare the problem size for LSS type II and III.  The $T=100$ KS equation solution was solved with $m=500$ time steps.  This means the KKT for LSS type II is 127128 by 127128.  For LSS type III, we used 25 time segments, so the linear system in equation \eqref{e:tanLSSsys} is 6223 by 6223, two orders of magnitude smaller than the type II KKT system.  However, it should be noted that convergence of MINRES was fairly slow for tangent and adjoint LSS.  The $T=100$ tangent and adjoint LSS solution converged to the specified relative residual L2 norm of $10^{-6}$ in $\sim5000$ iterations on average.  This large iteration count highlights the need for a highly efficient linear solver to implement LSS type III for large scale system like turbulent flow simulations.  As LSS type III is a boundary value problem in time, it is likely that solution methods based on the multigrid in space and time methods used by the authors for LSS type II will work well for LSS type III\cite{gomez:2013:masters,Blonigan:2013:MG}.  


\section{Conclusion}

In conclusion, unlike traditional sensitivity analysis methods, LSS can compute meaningful sensitivities for long-time-averaged quantities of interest in turbulent fluid flows, such as the average cumulative energy spectrum in homogeneous, isotropic turbulence. This result, along with the successful application of LSS to smaller problems such as the Lorenz equations and the Kuramoto-Sivashinsky equation shows the great potential of LSS as a new means to analyze chaotic and turbulent fluid flows.  

Two algorithms for LSS, type II and type III have been presented and demonstrated on turbulent flows and chaotic systems.  LSS type II solves the least squares problem by searching for an entire solution $v(t)$.  This results in a well conditioned problem with a very large KKT system.  LSS type III solves the least squares problem by searching for the solution at a number of checkpoints in time $v(t_0),v(t_1),...v(t_K)$.  This results in a reasonably well conditioned problem with a smaller KKT system than for type II.  However, convergence of LSS type III is very slow when solved with a Krylov solver (MINRES).  

Currently, work is being done to apply LSS type III to the homogeneous isotropic turbulence simulation shown in this paper and to a turbulent channel flow simulation.  Future work includes a study of how the time checkpoints used in LSS type III should be selected and what their effect on convergence is.  Also, the authors plan to explore using multigrid in time and space for LSS type III. Once we find an efficient solver, LSS will be used to investigate more complicated flows and geometries such as flow around a stalled airfoil or a turbulent jet in a cross flow.  

\section*{Acknowledgments}

The authors acknowledge AFOSR Award F11B-T06-0007 under Dr. Fariba Fahroo, NASA Award NNH11ZEA001N under Dr. Harold Atkins, as well as financial support from ConocoPhillips, the ANSYS fellowship, and the NDSEG fellowship.

\bibliography{main}
\bibliographystyle{aiaa}

\appendix

\section{Appendix: LSS type III}

\setlength\parindent{0pt}
\newcommand{\bB}{\uuline{\textbf{B}}}
\newcommand{\heta}{\hat{\eta}}
\newcommand{\evalt}[2]{\left. {#1} \right|_{{#2}}}
\newcommand{\intCHK}[1]{ \int_{t_i}^{t_{i+1}} {#1} \ dt}
\newcommand{\intCHKtau}[1]{ \int_{t_i}^{t_{i+1}} {#1} \ d\tau}

\subsection{Tangent Formulation}
\label{s:tanform}
Recall the least squares problem:
\begin{equation}
\min \frac12 \int_{t_0}^{t_K} v^T v \,dt \quad \mbox{s.t.}\quad
\frac{dv}{dt} - \frac{\partial f}{\partial u}\,v - \eta\,f =
\frac{\partial f}{\partial s}\;,
\end{equation}
We define $t_0$ as the run up time, $\Delta T_i = t_{i+1} - t_i,
i=0,\ldots,K-1$ be the length of the $i$th time segment, and $T \equiv t_K-t_0$.
For convenience, we denote
$\phi^{t,\tau}$ as the tangent propagator from $t$ to $\tau$, satisfying
\begin{equation}
\phi^{t,t} = I \mbox{ for all } t\;,\quad
\phi^{t',\tau}\cdot\phi^{t,t'} = \phi^{t,\tau}\;,\quad
\frac{d}{d\tau} \phi^{t,\tau} = \frac{\partial f}{\partial u}\bigg|_{\tau}
\phi^{t,\tau}\;,\quad
\frac{d}{dt} \phi^{*\,t,\tau} = -\frac{\partial f}{\partial u}\bigg|_{t}^*
\phi^{*\,t,\tau}\;,
\end{equation}
where star represents adjoint (or equivalently, transpose).  Because
\begin{equation}
\frac{df}{dt} = \frac{\partial f}{\partial u} f\;,\quad
f(\tau) = \phi^{t,\tau} f(t)\;.
\end{equation}
So a general solution to the linear constraint differential equation is
\begin{equation}
v(t) = \phi^{t_i,t} {\bf v}_{i}
+ \left(\int_{t_i}^{t} \eta_{i}(\tau)\,d\tau\right)f(t)
+ \int_{t_i}^{t} \phi^{\tau,t} \frac{\partial f}{\partial s}\bigg|_{\tau}\,d\tau\;,
\quad t_i\le t<t_{i+1}\;,
\label{e:tangent}
\end{equation}
where ${\bf v}_i = v(t_i)$.
The least squares problem becomes
\begin{equation}
\min \sum_{i=0}^{K-1} Q_i({\bf v}_i, s) \quad\mbox{s.t.}\quad
{\bf v}_{i+1} = \phi^{t_i,t_{i+1}} {\bf v}_{i}
+ \left(\int_{t_i}^{t_{i+1}} \eta_{i}(t)\,dt\right)f(t_{i+1})
+ \int_{t_i}^{t_{i+1}} \phi^{t,t_{i+1}}
  \frac{\partial f}{\partial s}\bigg|_t\,dt
\end{equation}
$Q_i$ is a quadratic form, whose variation is
\begin{equation}\begin{aligned}
   \delta Q_i &= \int_{t_i}^{t_{i+1}} v(t)^T \delta v(t) \,dt \\
&= \int_{t_i}^{t_{i+1}}
   v(t)^T
   \left(\phi^{t_i,t} \delta {\bf v}_i +
       \int_{t_i}^t\delta\eta_i(\tau) f(t)d\tau\right) dt \\
&= \int_{t_i}^{t_{i+1}}
   \left(\phi^{*\,t_i,t} v(t)\right)^T
   \delta {\bf v}_i +
   \left(\int_t^{t_{i+1}} v(\tau)^T f(\tau)\,d\tau \right)
   \delta\eta(t)\;dt
\end{aligned}\end{equation}
Note that we have exchanged integrals in the triangular area
$t_i<t<t_{i+1}, t_i<\tau<t_{i+1}$.
Therefore, the variation of the Lagrangian of the problem is
\begin{equation}\begin{aligned}
\delta \Lambda 
&= \sum_{i=0}^{K-1}
   \int_{t_i}^{t_{i+1}}
   \left(\phi_i^{*t} v(t)\right)^T
   \delta {\bf v}_i +
   \left(\int_t^{t_{i+1}} f(\tau)^T v(\tau)\,d\tau \right)
   \delta\eta_i(t)\;dt \\
&+ \sum_{i=1}^{K-1} \delta{\bf w}_i^T
   \left( {\bf v}_i - \phi^{t_{i-1},t_i}{\bf v}_{i-1}
        - \left(\int_{t_{i-1}}^{t_i} \eta(t)\,dt\right) f(t_i)
        - \int_{t_{i-1}}^{t_i} \phi^{t,t_i}
                \frac{\partial f}{\partial s}\bigg|_t\,dt\right) \\
&+ \sum_{i=1}^{K-1} {\bf w}_i^T
   \left( \delta {\bf v}_i - \phi^{t_{i-1},t_i}\delta{\bf v}_{i-1}
        - \left(\int_{t_{i-1}}^{t_i} \delta\eta(t)\,dt\right)f(t_i)
   \right)
\end{aligned}\end{equation}
where ${\bf w}_i,i=1,\ldots,K-1$ are Lagrange multipliers.
The variation must be 0, so all the collected terms in front of
$\delta{\bf v}_i$, $\delta{\bf w}_i$ and $\delta\eta_i$ must be 0.
\begin{equation} {\bf R}^v_i
 = \int_{t_i}^{t_{i+1}}\phi^{*\,t_i,t} v(t)\,dt
 + {\bf w}_i - \phi^{*\,t_i,t_{i+1}} {\bf w}_{i+1}
 = 0 \label{e:Rv} \end{equation}
\begin{equation} {\bf R}^w_i = {\bf v}_i - v(t_i^-) = 0 \label{e:Rw} \end{equation}
\begin{equation} R^{\eta}_i(t)
 = \int_t^{t_{i+1}}f(\tau)^T  v(\tau)\,d\tau
 - f(t_{i+1})^T{\bf w}_{i+1} = 0 
\end{equation}
for $i=0,\ldots,n-1$.
Here we have denoted ${\bf w}_0 = {\bf w}_K=0$,
The third equality gives
\begin{equation} f(t_i)^T {\bf w}_i = 0\;,\quad f(t)^T v(t) \equiv 0\;,\quad
f(t)^T w(t) \equiv 0 \label{e:tanProj} \end{equation}
Computationally, one can solve the tangent equation without $\eta$ for
\begin{equation}
v'(t) = v(t) - f(t)\,\int_{t_i}^{t} \eta(\tau)\,d\tau\;,
t_i\le t < t_{i+1}
\label{e:tanProjection}
\end{equation}
Then one can obtain $ v(t) = P_t\,v'(t) $ via a projection to the
orthogonal direction of $f(t)$ anytime.
Here
\begin{equation}
P_t\,v'(t) = v'(t) - \frac{v'(t)^{\top} f(t)}{f(t)^{\top} f(t)} f(t)
\label{e:Pdef}
\end{equation}
In addition, the component of $v'(t_{i+1}^-)$ along the
$f(t_{i+1})$ direction gives
\begin{equation} 
\Delta t_i\,\zeta_i = \int_{t_i}^{t_{i+1}} \eta_i(t)\,dt
\end{equation}
Then,
\begin{equation} \begin{aligned}
\int_{t_i}^{t_{i+1}} \frac{\partial J}{\partial u}\bigg|_{t}^{\top} v'\,dt
&= 
\int_{t_i}^{t_{i+1}} \frac{\partial J}{\partial u}\bigg|_{t}^{\top} v\,dt
- 
\int_{t_i}^{t_{i+1}} \int_{t_i}^{t} \frac{\partial J}{\partial u}\bigg|_{t}^{\top} f(t)
\eta(\tau)\,d\tau\,dt \\
&=
\int_{t_i}^{t_{i+1}} \frac{\partial J}{\partial u}\bigg|_{t}^{\top} v\,dt
- 
\int_{t_i}^{t_{i+1}} \eta(\tau)
\int_{\tau}^{t_{i+1}} \frac{\partial J}{\partial u}\bigg|_{t}^{\top} f(t) \,dt\,d\tau \\
&=
\int_{t_i}^{t_{i+1}} \frac{\partial J}{\partial u}\bigg|_{t}^{\top} v\,dt
- 
\int_{t_i}^{t_{i+1}} \eta(\tau) (J(t_{i+1}) - J(\tau)) \,d\tau \\
&=
\int_{t_i}^{t_{i+1}} \left(\frac{\partial J}{\partial u}\bigg|_{t}^{\top} v + \eta\,J\right)dt
- \Delta t_i\,\zeta_i\,J(t_{i+1})
\end{aligned} \end{equation}
Sensitivity of our long time averaged quantity of interest $\bar{J}$ can then be computed from the solution:
\begin{equation}
\frac{d\bar{J}}{ds} = 
\frac1T \int_{t_0}^{t_K} \left( \frac{\partial J}{\partial u}\bigg|_{t}^{\top} v + \eta \left(J - \bar{J} \right) \right) dt
= 
\frac1T \sum_{i=0}^{K-1}
\int_{t_i}^{t_{i+1}} \frac{\partial J}{\partial u}\bigg|_{t}^{\top} v'\ dt
+ \frac1T \sum_{i=0}^{K-1} \Delta t_i \zeta_i \left(J(t_{i+1}) - \bar{J}\right)
\label{e:tanGrad}
\end{equation}

Where $T \equiv t_K-t_0$.  

\subsection{Tangent Matrix System}
\label{s:tanmat}
\label{s:tanSys}

LSS type III can be thought of as iteratively solving the system:

\[
 \uuline{\bf A} \left[ \begin{array}{c}
    \bv_0 \\
    \bv_1 \\
    \vdots \\
    \bv_{K-1} \\
    \bw_1 \\
    \vdots \\
    \bw_{K-1}
   \end{array} \right] = \uline{\bf b}
\]

To determine $\uuline{\bf A}$ and $\uline{\bf b}$, we rearrange the adjoint equations \eqref{e:Rv} and \eqref{e:Rw}:

\begin{align}
&\int_{t_i}^{t_{i+1}} \prop{*\,t_i}{t} \left(\prop{t_i}{t} \bv_i + f(t) \int_{t_i}^t \eta(t) \ dt \right)  \ dt +  \bw_i - \prop{* \, t_i}{t_{i+1}}\bw_{i+1} \label{e:w} \\
&= -\int_{t_i}^{t_{i+1}} \prop{*\,t_i}{t} \int_{t_i}^t \prop{\tau}{t}\pd{f}{s}\bigg|_{\tau}\ d\tau \ dt, \qquad i = 0,1,...,K-1 \nonumber \\
&\bv_i -  \prop{t_{i-1}}{t_i} \bv_{i-1} - f(t_i) \int_{t_{i-1}}^{t_i} \eta(t) \ dt = \int_{t_{i-1}}^{t_{i}} \prop{t}{t_i}\pd{f}{s}\bigg|_{t}\ dt \qquad i = 1,2,...,K-1 \label{e:v}
\end{align}

To form $A$ and $b$ we need to rewrite \eqref{e:w} and \eqref{e:v} to eliminate any dependence on $\heta(t)$ while enforcing \eqref{e:tanProj}.  To do this, we first substitute equations \eqref{e:tanProj} and \eqref{e:tanProjection} into \eqref{e:v}:

\begin{equation}
 v(t_i) = P_{t_i}\prop{t_{i-1}}{t_i} \bv_{i-1} + P_{t_i}\int_{t_{i-1}}^t \prop{\tau}{t}\pd{f}{s}\bigg|_{\tau}\ d\tau
\label{e:vI}
\end{equation}

Where $P_{t_i}$ is the projection defined in \eqref{e:Pdef}.  Next, we can use \eqref{e:tanProj} to show that: 

\begin{equation}
\bw_i = P_{t_i}\prop{* \, t_i}{t_{i+1}}\bw_{i+1}  -P_{t_i}\int_{t_i}^{t_{i+1}} \prop{*\,t_i}{t}  v(t) \ dt  
\label{e:wI}
\end{equation}

Note that \eqref{e:vI} and \eqref{e:wI} are not dual consistent.  To make them consistent, we use the following identities, which are derived from \eqref{e:tanProjection}:  

\begin{align}
 \bv_i = P_{t_i} \bv_i \\
 \bw_i = P_{t_i} \bw_i 
\end{align}

Now, \eqref{e:w} and \eqref{e:v} can be rewritten as:

\begin{align}
&P_{t_i}\int_{t_i}^{t_{i+1}} \prop{*\,t_i}{t} P_t\prop{t_i}{t} P_{t_i} \bv_i  \ dt +  \bw_i - P_{t_i} \prop{* \, t_i}{t_{i+1}} P_{t_{i+1}}\bw_{i+1} \label{e:wII} \\
&= -P_{t_i}\int_{t_i}^{t_{i+1}} \prop{*\,t_i}{t} P_t \int_{t_i}^t \prop{\tau}{t}\pd{f}{s}\bigg|_{\tau}\ d\tau \ dt, \qquad i = 0,1,...,K-1 \nonumber \\
&\bv_i -  P_{t_i} \prop{t_{i-1}}{t_i} P_{t_{i-1}} \bv_{i-1} = P_{t_i} \int_{t_{i-1}}^{t_{i}} \prop{t}{t_{i}}\pd{f}{s}\bigg|_{t}\ dt \qquad i = 1,2,...,K-1 \label{e:vII}
\end{align}


Finally, we can derive an expression for the matrix system.  First we define a few $n\times n$ matrices, where $n$ is the dimension of the primal solution $u(t)$:

\begin{equation}
 D_i \equiv \int_{t_i}^{t_{i+1}} P_{t_i}\prop{*\,t_i}{t} P_t\prop{t_i}{t} P_{t_i}\ dt, \qquad F_i \equiv P_{t_{i+1}} \prop{t_{i}}{t_{i+1}} P_{t_{i}}
\end{equation}

It can be shown that $P_{t_i} = I - \frac{1}{f(t_i)^Tf(t_i)}f(t_i)f(t_i)^T$, which is a symmetric matrix.  Therefore, the product $\prop{*\,t_i}{t} P_t\prop{t_i}{t}$ is a symmetric matrix and $D_i$ is symmetric.  On the other hand, $F_i$ is not necessarily symmetric, as $\prop{t_{i}}{t_{i+1}}$ is not necessarily symmetric.

We also define two $n \times 1$ vectors which appear in the right hand side $\uline{\bf b}$:

\begin{equation}
 a_i \equiv -P_{t_i}\int_{t_i}^{t_{i+1}} \prop{*\,t_i}{t} P_t \int_{t_i}^t \prop{\tau}{t}\pd{f}{s}\bigg|_{\tau}\ d\tau \ dt, \qquad b_i \equiv P_{t_i} \int_{t_{i-1}}^{t_{i}} \prop{t}{t_{i}}\pd{f}{s}\bigg|_{t}\ dt
\end{equation}

Using the above definitions, the matrix system is:

\begin{equation}
\left( \begin{array}{ccccc|cccc}
D_0 & & & & &  F_0^T & & & \\
& D_1 & & & &  I & F_1^T & & \\
& & \ddots &  & & & I & \ddots & \\
 & & & D_{K-2} & &    & & \ddots & F_{K-1}^T \\
& & & & D_{K-1} & & & & I \\\hline
 F_0 & I & &  & & & & & \\
 & F_1 & I & &  & & & & \\
 & & \ddots & \ddots & & & & & \\
 & & & F_{K-1} & I & & & & \end{array} \right) \left(\begin{array}{c} 
\bv_0\\
\bv_1 \\
\vdots \\
\vdots \\
 \bv_{K-1} \\\hline
\bw_1 \\
 \bw_2 \\
\vdots \\
\bw_{K-1} \end{array} \right) = \left(\begin{array}{c} 
a_0 \\
a_1 \\
\vdots \\
\vdots \\
a_{K-1} \\\hline
b_1\\
b_2 \\
\vdots \\
b_{K-1} \end{array} \right)
\label{e:tanSys}
\end{equation}

The upper partition corresponds to equation \eqref{e:wII} and the lower partition corresponds to \eqref{e:vII}.  As the $D_i$ block matrices are symmetric, the overall matrix, $\uuline{\bf A}$ is symmetric.  However, by the definition of $P_{t_i}$, each $D_i$ and $F_i$ block is a full matrix.

\subsection{Adjoint Formulation}
\label{s:adjform}

To derive the discrete Adjoint system of equation \eqref{e:tanSys}, we first need to derive an expression for the gradient in terms of $\bv_i$ and $\bw_i$.  First, we rewrite \eqref{e:tanGrad} as: 

\begin{equation}
\dd{\bar{J}}{s} = \frac{1}{T}\sum_{i=0}^{K-1} \left( \intCHK{\evalt{\pd{J}{u}}{t}^{\top} v'} + \Delta t_i \zeta_i (J_{i+1} - \bar{J} )\right)
\label{e:tan}
\end{equation}

where $J_{i+1} \equiv J(t_{i+1})$.  It can be shown that:

\begin{equation}
\Delta t_i \zeta_i = \int_{t_i}^{t_{i+1}} \eta(\tau) d\tau = -\frac{f_{i+1}^T v'(t_{i+1})}{f_{i+1}^T f_{i+1}}
\label{e:zeta}
\end{equation}

where $f_{i+1} \equiv f(t_{i+1})$.  Also, equations \eqref{e:tangent} and \eqref{e:tanProjection} can be combined to obtain:

\begin{equation}
v'(t) = \prop{t_i}{t} P_{t_i} \bv_i + \int_{t_i}^t \prop{\tau}{t} \evalt{\pd{f}{s}}{\tau}d\tau
\label{e:vprime}
\end{equation}

The identity $P_{t_i} \bv_i$ is added to make \eqref{e:vprime} consistent with \eqref{e:vII}.  Next, we combine equations \eqref{e:tan}, \eqref{e:zeta}, and \eqref{e:vprime}:

\begin{align*}
\dd{\bar{J}}{s} &= \frac{1}{T} \sum_{i=0}^{K-1}  \intCHK{\evalt{\pd{J}{u}}{t}^{\top}  \left(\prop{t_i}{t} P_{t_i} \bv_i + \int_{t_i}^t \prop{\tau}{t} \pd{f}{s}d\tau \right) }\\ &- \frac{(J_{i+1} - \bar{J})}{f_{i+1}^T f_{i+1}} f_{i+1}^T \left(\prop{t_i}{t_{i+1}} P_{t_i} \bv_i + \int_{t_i}^{t_{i+1}} \prop{\tau}{t_{i+1}} \pd{f}{s}d\tau \right)
\end{align*}

This can be reorganized into terms dependent on $\bv_i$ and $\pd{f}{s}$:

\begin{align*}
\dd{\bar{J}}{s} &= \frac1T\sum_{i=0}^{K-1}  \left[ \intCHK{  \evalt{\pd{J}{u}}{t}^{\top} \prop{t_i}{t}} - \frac{(J_{i+1} - \bar{J})}{ f_{i+1}^T f_{i+1}} f_{i+1}^T \prop{t_i}{t_{i+1}} \right] P_{t_i} \bv_i   \\ &+  \left[ \intCHK{ \evalt{\pd{J}{u}}{t}^{\top}  \int_{t_i}^{t}\prop{\tau}{t}\evalt{\pd{f}{s}}{\tau}d\tau } - \frac{(J_{i+1} - \bar{J})}{ f_{i+1}^T f_{i+1}} \int_{t_i}^{t_{i+1}} f_{i+1}^T \phi_i^{\tau,t_{i+1}} \evalt{\pd{f}{s}}{\tau}d\tau \right] 
\end{align*}

We can rearrange the order of integration in the terms dependent on $\pd{f}{s}$ to obtain:

\begin{align*}
\dd{\bar{J}}{s} &= \frac1T\sum_{i=0}^{K-1}  \left[ \intCHK{  \evalt{\pd{J}{u}}{t}^{\top}  \prop{t_i}{t}} - \frac{(J_{i+1} - \bar{J})}{ f_{i+1}^T f_{i+1}} f_{i+1}^T \prop{t_i}{t_{i+1}} \right] P_{t_i} \bv_i   \\ &+  \left[\intCHKtau{ \left( \int_{\tau}^{t_{i+1}} \evalt{\pd{J}{u}}{t}^{\top} \prop{\tau}{t}  dt \right) \evalt{\pd{f}{s}}{\tau}} - \frac{(J_{i+1} - \bar{J})}{ f_{i+1}^T f_{i+1}} \int_{t_i}^{t_{i+1}} f_{i+1}^T \prop{\tau}{t_{i+1}} \evalt{\pd{f}{s}}{\tau}d\tau \right] 
\end{align*}

Now, we write the gradient in terms of adjoint propagators $\prop{*\tau}{t}$:

\begin{align*}
\dd{\bar{J}}{s} &= \frac1T \sum_{i=0}^{K-1}  \left[ \intCHK{  \left(\prop{*t_i}{t} \evalt{\pd{J}{u}}{t} \right)^T} - \frac{(J_{i+1} - \bar{J})}{ f_{i+1}^T f_{i+1}} \left( \prop{*t_i}{t_{i+1}} f_{i+1}\right)^T \right] P_{t_i} \bv_i  \\ &+  \left[\intCHKtau{ \left( \int_{\tau}^{t_{i+1}} \left( \prop{*\tau}{t} \evalt{\pd{J}{u}}{t}  \right)^T  dt \right) \evalt{\pd{f}{s}}{\tau}} - \frac{(J_{i+1} - \bar{J})}{ f_{i+1}^T f_{i+1}} \int_{t_i}^{t_{i+1}} \left( \prop{*\tau}{t_{i+1}} f_{i+1} \right)^T \evalt{\pd{f}{s}}{\tau} d\tau \right] 
\end{align*}

Define:

\[
g_i(\tau) = \frac1T \int_{\tau}^{t_{i+1}}  \prop{*\tau}{t} \evalt{\pd{J}{u}}{t}  \  dt - \frac{(J_{i+1} - \bar{J})}{Tf_{i+1}^T f_{i+1}} \prop{*\tau}{t_{i+1}} f_{i+1}  
\]

Using this definition, equation \eqref{e:tan} becomes:

\[
\dd{\bar{J}}{s} = \sum_{i=0}^{K-1} \left(  g_i^T(t_i) P_{t_i} \bv_i   +  \intCHKtau{g_i^T(\tau) \evalt{\pd{f}{s}}{\tau} } \right)
\]

Define the vector $\uline{g}$, where $\uline{g}_i = P_{t_i} g_i(t_i) $.  Also define:

\[
\pd{\bar{J}}{s} \equiv \sum_{i=0}^{K-1} \intCHKtau{g_i^T(\tau) \evalt{\pd{f}{s}}{\tau} }
\]

Define $\uline{\bv}$ and $\uline{\bw}$ as vectors containing $\bv_i$ for $i = 0,1,...,K-1$ and $\bw_i$ for $i = 1,...,K-1$, respectively.  Also define $\uline{a}$ and $\uline{b}$ as vectors that contain $a_i$ (see appendix \ref{s:tanSys}) for $i = 0,1,...,K-1$ and $b_i$ for $i = 1,...,K-1$, respectively.  We can now write discrete equations to compute the gradient:

\begin{equation}
 \uuline{\bf A} \left(\begin{array}{c}
                 \uline{\bv} \\
		 \uline{\bw}
                \end{array}\right) = \uline{\bf b} = \left(\begin{array}{c}
                 \uline{a} \\
		 \uline{b}
                \end{array}\right), \qquad \dd{\bar{J}}{s} = \left(\uline{g}^T \quad \uline{0}^T \right) \left(\begin{array}{c}
                 \uline{\bv} \\
		 \uline{\bw}
                \end{array}\right) + \pd{\bar{J}}{s}
\end{equation}

As $ \uuline{\bf A}$ is symmetric, the adjoint equations are:

\begin{equation}
 \uuline{\bf A} \left(\begin{array}{c}
                 \uline{\hbv} \\
		 \uline{\hbw}
                \end{array}\right)  = \left(\begin{array}{c}
                 \uline{g} \\
		 \uline{0}
                \end{array}\right), \qquad \dd{\bar{J}}{s} = \left(\uline{a}^T \quad \uline{b}^T \right) \left(\begin{array}{c}
                 \uline{\hbv} \\
		 \uline{\hbw}
                \end{array}\right) + \pd{\bar{J}}{s}
\label{e:discAdj}
\end{equation}

Where $\uline{\hbv}$ and $\uline{\hbw}$ contain the adjoint variables $\hbv_i$ and $\hbw_i$.  

From \eqref{e:discAdj}, we find that the adjoint equation is:

\begin{align}
&P_{t_i}\int_{t_i}^{t_{i+1}} \prop{*\,t_i}{t} P_t\prop{t_i}{t} P_{t_i} \hbv_i  \ dt +  \hbw_i - P_{t_i} \prop{* \, t_i}{t_{i+1}} P_{t_{i+1}}\hbw_{i+1} \label{e:wAdj} \\
&= P_{t_i} \intCHK{  \prop{*t_i}{t} \evalt{\pd{J}{u}}{t} } - P_{t_i}\frac{(J_{i+1} - \bar{J})}{ f_{i+1}^T f_{i+1}} \prop{*t_i}{t_{i+1}} f_{i+1}, \qquad i = 0,1,...,K-1 \nonumber \\
&\hbv_i -  P_{t_i} \prop{t_{i-1}}{t_i} P_{t_{i-1}} \hbv_{i-1} = 0 \qquad i = 1,2,...,K-1 \label{e:vAdj}
\end{align}

Finally, an expression for $\dd{\bar{J}}{s} $ can be derived from \eqref{e:discAdj}:  

\begin{align}
 \dd{\bar{J}}{s} &= \sum_{i=0}^{K-1} \left( -P_{t_i}\int_{t_i}^{t_{i+1}} \prop{*\,t_i}{t} P_t \int_{t_i}^t \prop{\tau}{t}\evalt{\pd{f}{s}}{\tau}\ d\tau \ dt \right)^T \hbv_i \nonumber\\
&  + \sum_{i=0}^{K-1} \left( P_{t_{i+1}} \int_{t_{i}}^{t_{i+1}} \prop{t}{t_{i+1}}\evalt{\pd{f}{s}}{t}\ dt \right)^T \hbw_{i+1} \\ 
& + \sum_{i=0}^{K-1} \intCHKtau{ \evalt{\pd{f}{s}}{\tau}^T \left( \int_{\tau}^{t_{i+1}} \prop{*\tau}{t} \evalt{\pd{J}{u}}{t}   dt - \frac{(J_{i+1} - \bar{J})}{ f_{i+1}^T f_{i+1}}  \prop{*\tau}{t_{i+1}} f_{i+1} \right) } \nonumber 
\end{align}

Note that $P_{t_K}\hbw_K = 0$ for the above expression to hold. Next, we can rearrange the order of integration in the first term: 

\begin{align}
 \dd{\bar{J}}{s} &= \sum_{i=0}^{K-1} \left( -P_{t_i}\int_{t_i}^{t_{i+1}} \int_{\tau}^{t_{i+1}} \prop{*\,t_i}{t} P_t \prop{\tau}{t}\evalt{\pd{f}{s}}{\tau}\ dt \ d\tau \right)^T \hbv_i \nonumber\\
&  + \sum_{i=0}^{K-1} \int_{t_{i}}^{t_{i+1}} \evalt{\pd{f}{s}}{t}^T \left( \prop{* \, t}{t_{i+1}} P_{t_{i+1}}\hbw_{i+1} \right) \ dt  \\ 
& + \sum_{i=0}^{K-1} \intCHKtau{ \evalt{\pd{f}{s}}{\tau}^T \left( \int_{\tau}^{t_{i+1}} \prop{*\tau}{t} \evalt{\pd{J}{u}}{t}   dt - \frac{(J_{i+1} - \bar{J})}{ f_{i+1}^T f_{i+1}}  \prop{*\tau}{t_{i+1}} f_{i+1} \right) } \nonumber 
\end{align}

The first term can now be rewritten and combined with the second term:

\begin{align}
 \dd{\bar{J}}{s} &= \sum_{i=0}^{K-1} \int_{t_i}^{t_{i+1}} \evalt{\pd{f}{s}}{\tau}^T \left( -\int_{\tau}^{t_{i+1}} \prop{*\,\tau}{t} P_t \prop{t_i}{t} P_{t_i} \hbv_i \ dt + \prop{* \, \tau}{t_{i+1}} P_{t_{i+1}}\hbw_{i+1} \right) \ d\tau \label{e:AdjGradI} \\ 
& + \sum_{i=0}^{K-1} \intCHKtau{ \evalt{\pd{f}{s}}{\tau}^T \left( \int_{\tau}^{t_{i+1}} \prop{*\tau}{t} \evalt{\pd{J}{u}}{t}   dt - \frac{(J_{i+1} - \bar{J})}{ f_{i+1}^T f_{i+1}}  \prop{*\tau}{t_{i+1}} f_{i+1} \right) } \nonumber 
\end{align}

From \eqref{e:wAdj} and \eqref{e:vAdj}, it can be seen that:

\begin{align}
 \hw(t) =& - \int_{t}^{t_{i+1}} \prop{*\,t_i}{t} P_t \hv'(t)  \ dt + \prop{* \, t}{t_{i+1}} P_{t_{i+1}}\hbw_{i+1} \\ &+ \intCHK{  \prop{*t_i}{t} \evalt{\pd{J}{u}}{t} } - \frac{(J_{i+1} - \bar{J})}{ f_{i+1}^T f_{i+1}} \prop{*t}{t_{i+1}} f_{i+1} \nonumber
\end{align}

Therefore \eqref{e:AdjGradI} can be rewritten as:

\begin{equation}
 \dd{\bar{J}}{s} = \sum_{i=0}^{K-1} \int_{t_i}^{t_{i+1}} \evalt{\pd{f}{s}}{t}^T \hw(t) \ dt 
\label{e:AdjGradII}
\end{equation}

\end{document}